\documentclass[amsmath,amssymb,11pt]{revtex4}

\usepackage{graphicx}
\usepackage{amssymb}
\usepackage{epstopdf}
\usepackage{epsfig}
\allowdisplaybreaks

\begin{document}

\def\aprge{\buildrel > \over {_{\sim}}}
\def\aprle{\buildrel < \over {_{\sim}}}

\def\etal{{\it et.~al.}}
\def\ie{{\it i.e.}}
\def\eg{{\it e.g.}}

\def\bwt{\begin{widetext}}
\def\ewt{\end{widetext}}
\def\be{\begin{equation}}
\def\ee{\end{equation}}
\def\bea{\begin{eqnarray}}
\def\eea{\end{eqnarray}}
\def\bean{\begin{eqnarray*}}
\def\eean{\end{eqnarray*}}
\def\bary{\begin{array}}
\def\eary{\end{array}}
\def\bi{\bibitem}
\def\bit{\begin{itemize}}
\def\eit{\end{itemize}}

\def\lan{\langle}
\def\ran{\rangle}
\def\lra{\leftrightarrow}
\def\la{\leftarrow}
\def\ra{\rightarrow}
\def\dash{\mbox{-}}
\def\ol{\overline}

\def\ub{\ol{u}}
\def\db{\ol{d}}
\def\sb{\ol{s}}
\def\cb{\ol{c}}

\def\re{\rm Re}
\def\im{\rm Im}

\def \b{{\cal B}}
\def \ca{{\cal A}}
\def \ko{K^0}
\def \ok{\overline{K}^0}
\def \s{\sqrt{2}}
\def \st{\sqrt{3}}
\def \sx{\sqrt{6}}
\title{{\bf Properties of the Scalar Mesons below 1.0 GeV as Hadronic Molecules}}
\author{Yong-Liang Ma}
\address{Institute of High Energy Physics, CAS, P.O.Box 918(4), Beijing 100049, China\\
Theoretical Physics Center for Science Facilities, CAS, Beijing
100049}
\date{\today}
\begin{abstract}
Properties of scalar mesons below 1.0 GeV have been studied by
regarding them as hadronic molecular states. Using the effective
Lagrangian approach, we have calculated their leptonic decays,
strong decays and productions via the $\phi$ meson radiative decays.
Comparing our results with that given in the literature and the
data, we conclude that it is difficult to arrange them in the same
nonet if some of them are pure hadronic molecules.
\end{abstract}
\maketitle

\section{Introduction}

\label{Int}

Understanding the nature of scalar mesons is a prominent topic in
the past 30-40 years. The importance of the nature of the scalar
mesons is that, because the properties of the scalar mesons,
especially that with masses below 1.0 GeV, are difficult to be
understood in the constituent quark models, the study of the light
scalar mesons can help us to understand the nonperturbative
properties of QCD. Moreover, because the scalar mesons have the same
quantum numbers as the vacuum, it can help us to reveal the
mechanism of symmetry breaking which is, up to now, one of the most
profound problems in particle physics.

Many properties of scalar mesons are not so clear although they have
been investigated for several decades. Experimentally, it is
difficult to identify the scalar mesons because of their large decay
widths which cause strong overlaps between resonances and grounds
and also because several decay channels open up within a short mass
interval. And due to these problems many data on scalar mesons are
not so precise, so that it is not so easy to reveal their underlying
structures~\cite{Amsler:2008zz,Godfrey:1998pd,Close:2002zu}.
Theoretically, there are many calculations based on different
models, but it seems that one cannot rule out some of them based on
the present measurements. Currently, the observation shows that the
known $0^{++}$ mesons below 2.0 GeV can be classified into to
classes: One class with masses below (or near) 1.0 GeV and the other
class with masses above 1.0 GeV~\cite{Amsler:2008zz}.

To study light scalar mesons, one problem we have to confront is how
to classify the present observed scalar objects. One opinion is that
the scalar objects with masses below 1.0 GeV, including two
isosinglets $\sigma(600)$ and $f_0(980)$, one isotriplet $a_0(980)$
and two isodoublets $K_0^{\ast}(800)$, can be classified into one
nonet~\cite{Jaffe:1976ig,Jaffe:1976ih,Oller:1998hw,Ishida:1999qk,Alford:2000mm,Maiani:2004uc,Scadron:2003yg}.
On the contrary, another opinion, inspired by linear sigma model and
unitary quark model, is that $\sigma(600), f_0(980), a_0(980)$ and
$K_0^{\ast}(1430)$ form a scalar
nonet~\cite{Tornqvist:1995kr,Tornqvist:1999tn}.

Concerning the structures of light scalar mesons, they are still
open questions so far although many attempts have been made to
understand them in the literature. For the light scalars mesons with
masses below 1.0 GeV, some people, by considering their dominant
two-body decays, believe that they are multiquark (or multiquark
dominant) states
~\cite{Jaffe:1976ig,Jaffe:1976ih,Weinstein:1990gu,Alford:2000mm,Maiani:2004uc,Pelaez:2004xp,Pelaez:2003dy,Pelaez:2006nj}
or hadronic molecular
states~\cite{Weinstein:1990gu,Giacosa:2007bs,Branz:2007xp,Branz:2008qm,Branz:2008ha,Branz:2008cb}
( Note that some references list here do not different multiquak
state from molecular state.). Alternatively, properties of some of
these light scalar mesons were also investigated in the $q\bar{q}$
picture~\cite{Tornqvist:1995kr,Faessler:2003yf,vanBeveren:2001kf,Celenza:2000uk}.
Moreover, in Ref.~\cite{Dai:2003ip}, the spectrum of light scalar
mesons below 1.0 GeV were studied in the $q\bar{q}$ picture with
including the instanton effect. And it was found that the $q\bar{q}$
components with the instanton or tetraquark effect can explain the
spectrum at a certain level. In Ref.~\cite{:2008my}, we also studied
the the effect of instanton-induced interaction in light meson
spectrum on the basis of the phenomenological harmonic models for
quarks. For the light scalar mesons with masses above 1.0 GeV, the
potential model calculation indicates that they are $q\bar{q}$
states. However, concerning the spectrum of the scalar mesons with
masses above 1.0 GeV, since there are three isoscalar states,
$f_1(1370), f_0(1500)$ and $f_0(1710)$, one believes that there is a
glueball candidate among them. Based on the recent lattice results,
the mixing between glueball and $q\bar{q}$ components in these three
objects were fit in Ref.~\cite{Cheng:2009dg}. It should be noted
that, the two problems we mentioned above are not independent. If we
know the classification of the scalar mesons and the structures of
some states, we may deduce the structures of other states according
to the classification, and inversely, if the structures of the
scalar mesons are confirmed the classification becomes obvious.

In this paper, since some scalar mesons (mainly $f_0(980)$ and
$a_0(980)$) have been studied in the literature and the numerical
results yielded there are consistent with the data, we will study
the properties of all the scalar mesons with masses below 1.0 GeV by
regarding all of them as pure hadronic molecules and check whether
they form one nonet in the effective Lagrangian approach. Our logic
is, if the scalar meson with masses below 1.0 GeV form a nonet, they
should have the same structures. And if some of them can be
interpreted as hadronic molecules as discussed in the
literature~\cite{Giacosa:2007bs,Branz:2007xp,Branz:2008qm,Branz:2008ha,Branz:2008cb},
all the elements in the same nonet should be interpreted as hadronic
molecules. If our start point is reasonable, the yielded results
should be consistent with the data. Otherwise, it is difficult to
classify them into the same nonet, at least in the the hadronic
molecular interpretation.

In the molecular picture the interaction of scalar meson to its
constituents can be described by the effective Lagrangian. The
corresponding effective coupling constant $g_{_{S}}$ is determined
by the compositeness condition $Z = 0$ which was earlier used by
nuclear physicists and is being widely used by particle
physicists(see the references in \cite{Faessler:2007gv}). In
Refs.~\cite{Faessler:2007gv,Faessler:2007us,Dong:2008gb} this method
has been applied to study the newly observed charmed mesons
\cite{Faessler:2007gv,Faessler:2007us,Dong:2008gb} and the decay
properties calculated there are consistent with the observed data.
We had employed the above technique to predict the decay properties
of the bottom-strange mesons\cite{Faessler:2008vc} and recently we
applied this method to studied the baryonium picture of
$X(1835)$~\cite{Ma:2008hc}. The production and decay properties of
some scalar mesons have been studied in
Refs.~\cite{Giacosa:2007bs,Branz:2007xp,Branz:2008qm,Branz:2008ha,Branz:2008cb}
with the help of this technique by regarding them as pure hadronic
molecules and the numerical results yielded there agree with the
data.

It should be mentioned that, in the tetraquark interpretation of the
scalar mesons with masses below 1.0 GeV, the diquark $(qq)$ is in
the $\underline{3}$ representation $(us,ds,ud)$ of $SU(3)$ group so
there are at most two valence strange quarks in the $\sigma(600)$
and $f_0(980)$ as shown in Fig.~\ref{fig:nonet}. However in the
hadronic molecular interpretation, an isosinglet $\eta$ constituent
in addition to the isodoublet $(K^+,K^0)$ should be introduced to
form a nonet. So that there is $\eta\eta$ component, or
$(s\bar{s})(s\bar{s})$ at the quark level, in $\sigma(600)$ and
$f_0(980)$ which makes it is difficult to understand the scalar
meson spectrum from the quark level. Fortunately, in our present
work we adopt the scalar meson spectrum as input, with the help of
compositeness condition, to determine the coupling constant $g_{_S}$
between the molecule and its constituents.

In our explicit calculation of the effective coupling constant
$g_{_S}$, we will consider two cases, the one is local interaction
case and the other one is nonlocal interaction case with including a
correlation function which describes the distribution of the
constituents among the molecules. As in our previous works, we will
introduce a typical scale parameter $\Lambda_S$ to describe the
finite size of the molecules. The numerical results show, in the
region where the coupling constant $g_{_{S}}$ is stable against
$\Lambda_S$, the yielded value of $g_{_{S}}$ for the corresponding
molecular scalar meson, is consistent with that yielded from the
local interaction picture. So that in the other calculations we will
take the local interaction vertex. For other interactions, we will
resort to the phenomenological Lagrangian and borrow the relevant
coupling constants from the existing literature.

To determine the mixing angle $\theta_S$ between $\sigma(600)$ and
$f_0(980)$, we will apply the data for $f_0(980) \to 2\gamma$. It is
natural that only with this data one cannot fix $\theta_S$ uniquely
and to fix $\theta_S$ a well measured process of $\sigma(600)$, for
example $\sigma(600) \to 2\gamma$, is necessary. Concerning that the
coupling constant $g_{\sigma}$ depends on the mass of $\sigma(600)$
closely, we will not use this data, and will not discuss the physics
of $\sigma(600)$ in the present work.

With the determined coupling constant $g_{_S}$ we will calculate
leptonic decay constants of the scalar mesons and the relevant
leptonic decay widths. And from the yielded numerical results, we
find that the leptonic decay widths of the scalar mesons are too
small to be measured at present. With our hadronic interpretation,
the strong decays of the scalar mesons to two-pseudoscalar mesons
have also been calculated. The numerical results for the strong
decays indicate that it is difficult to arrange the scalar mesons
with masses below 1.0 GeV into the same nonet in the hadronic
molecular interpretation. To study the productions of the scalar
mesons $f_0(980)$ and $a_0(980)$ in the radiative decays of $\phi$
meson, we will includ the final state interaction effect due to the
two axial-vector mesons nonets with quantum numbers $J^{PC} =
1^{++}$ and $1^{+-}$. We find the final state interaction plays a
negligible role in the production rates.

This paper is organized as the following: In section~\ref{sec:mol},
we discuss the molecular structures of the light scalar nonet with
masses below 1.0 GeV and calculate the coupling constant $g_{_{S}}$,
the leptonic decay constants and the leptonic decay widths. In
section~\ref{sec:strong}, the strong decays of the scalar mesons are
calculated in the framework of effective Lagrangian approach. The
productions of the $f_0$ and $a_0$ mesons in the radiative decays of
$\phi$ meson are studied in section~\ref{sec:radiativephi}. Our
discussions and conclusions are given in the last section.

\section{Molecular Structures of the Scalar Meson Nonet below 1.0 GeV}

\label{sec:mol}

In this section, we will discuss the properties of the scalar meson
nonet below 1.0 GeV in the hadronic molecular explanation, i.e.,
regarding them as two-pseudoscalar-meson bound states. At first, we
will identify pseudoscalar meson contents of the scalar mesons and
construct the effective Lagrangian describing the interaction
between scalar meson and its constituents. Then the magnitude of the
coupling constant $g_{_{S}}$ is calculated with the help of
compositeness condition and the leptonic decay constants and the
leptonic decay widths of the scalar mesons are yielded by the
standard loop integral.

\subsection{Molecular Structures of the Scalar Nonet}

\begin{figure}[htbp]
\begin{center}
\includegraphics[scale=0.4]{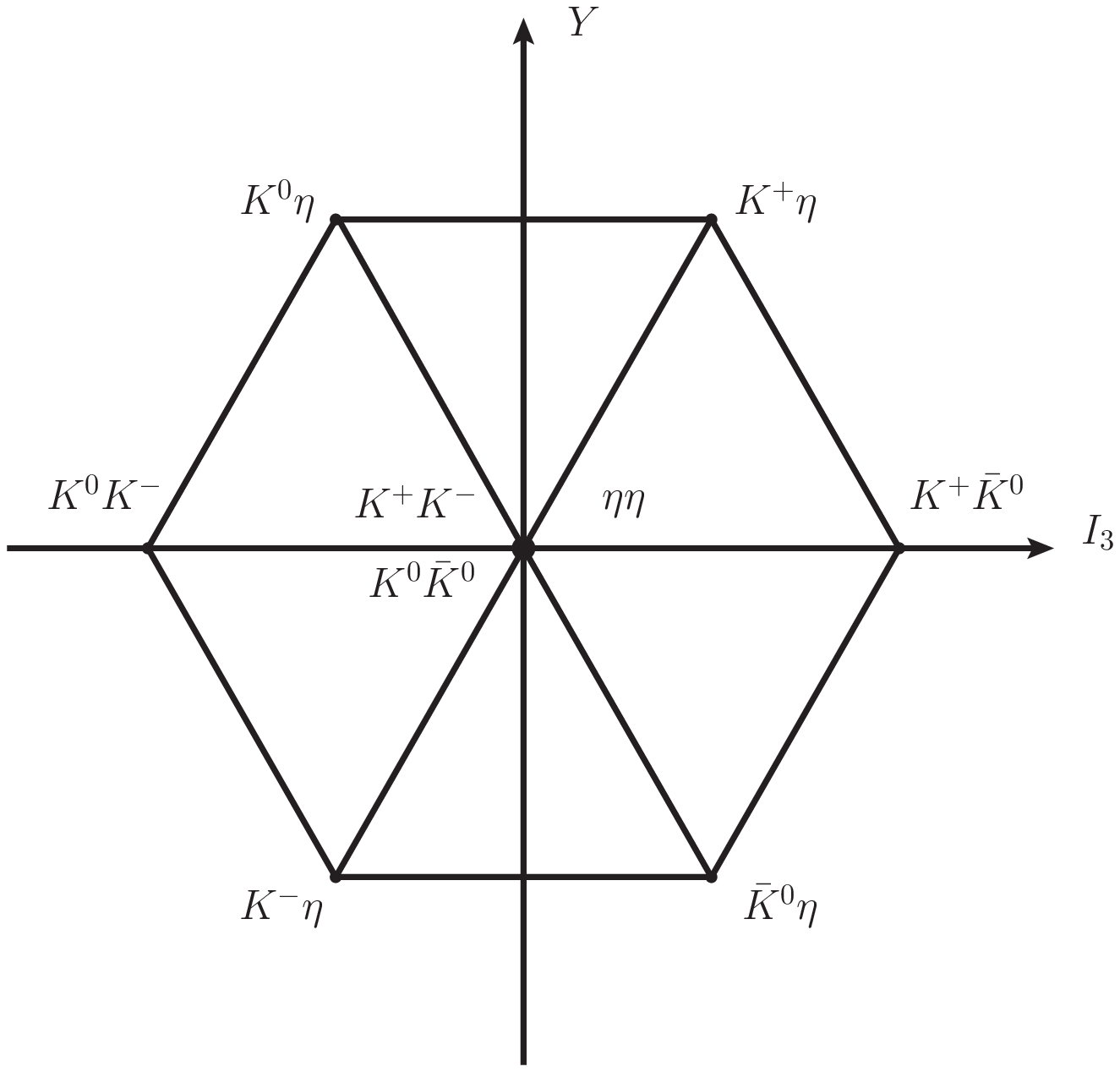}\includegraphics[scale=0.4]{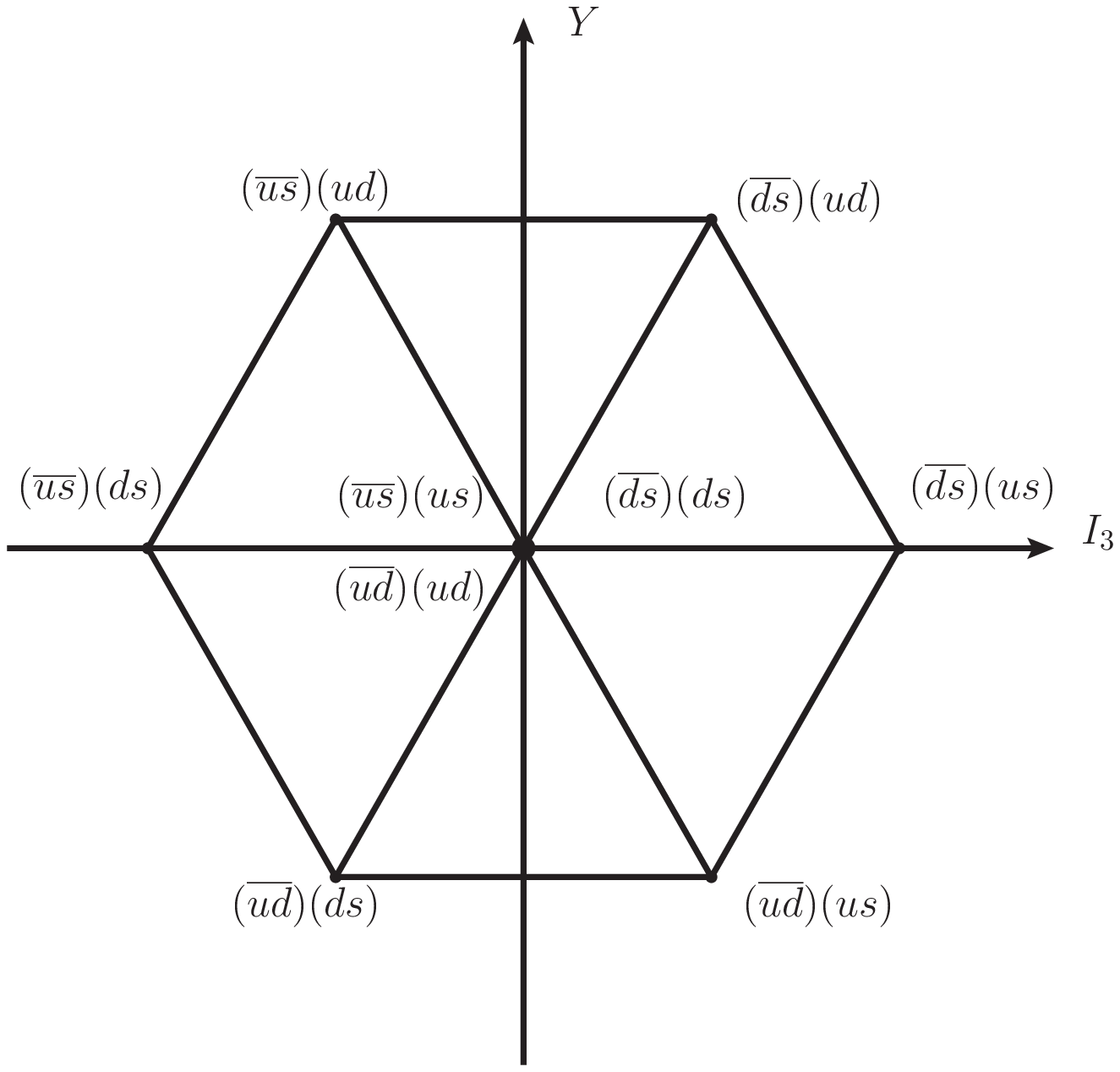}
\end{center}
\caption[Hadronic molecules (left) and tetraquark explanations (right) of scalar mesons in the $(I_3,Y)$ plane.]{%
Hadronic molecules (left) and tetraquark explanations (right) of
scalar mesons in the $(I_3,Y)$ plane. } \label{fig:nonet}
\end{figure}

Now, we are in the position to identify the constituents of light
scalar meson nonet as hadronic molecules. As we pointed above that
we will accept the opinion that the light scalar mesons: two
isosinglets $\sigma(600)$ and $f_0(980)$, one isotriplet $a_0(980)$
and two isodoublets $K_0^{\ast}(800)$, can be classified into one
nonet, we should have three elements at least to form this nonet.
And considering the physical requirement that the bound state should
have a mass smaller than the threshold of the corresponding
constituents, the constituents of the light scalar mesons should be
$K^+, K^0$ and $\eta$. With these considerations, we may arrange the
two-pseudoscalar-meson bound states in the $(I_3,Y)$ plane (where
$Y$ is the hypercharge which is relative to the strangeness $S$ via
relation $Y = S$ for mesons and relates the charge $Q$ and the third
component of isospin $I_3$ via relation $Q = I_3 + (1/2)Y$) as
Fig.~\ref{fig:nonet}. As a comparison, we also illustrate the quark
contents of the scalar nonet in the tetraquark interpretation in
Fig.~\ref{fig:nonet}.

Comparing with the quantum numbers of the light scalar mesons, we
write down the hadronic molecular structures of two isodoublets and
isotriplet as
\begin{eqnarray}
\left(
  \begin{array}{c}
    | K_0^{\ast +}(800)\rangle \\
    | K_0^{\ast 0}(800)\rangle \\
  \end{array}
\right)
 & = & \eta \left(
         \begin{array}{c}
           K^+ \\
           K^0 \\
         \end{array}
         \right);  \;\;\;\;\;
 \left(
   \begin{array}{c}
     | \bar{K}_0^{\ast 0}(800)\rangle \\
     | K_0^{\ast -}(800)\rangle \\
   \end{array}
 \right)
 =  \eta \left(
        \begin{array}{c}
          \bar{K}^0 \\
          K^- \\
        \end{array}
      \right) \nonumber\\
\left(
   \begin{array}{c}
     | a_0^+(980)\rangle \\
     | a_0^-(980)\rangle \\
   \end{array}
 \right) & = & \left(
        \begin{array}{c}
          K^+\bar{K}^0 \\
          K^0K^- \\
        \end{array}
      \right) ;  \;\;\;\;\;
| a_0^0(980)\rangle  = \frac{1}{\sqrt{2}}(| K^+ K^- \rangle - |
\bar{K}^0 K^0 \rangle )
\end{eqnarray}
and two singlets as
\begin{eqnarray}
S_8 & = & \frac{1}{\sqrt{6}} ( | K^+ K^-\rangle + | K^0\bar{K}^0
\rangle - 2 | \eta\eta \rangle ) \nonumber\\
S_0 & = & \frac{1}{\sqrt{3}} ( | K^+ K^-\rangle + | K^0\bar{K}^0
\rangle + | \eta\eta \rangle )
\end{eqnarray}
The physical states $\sigma(600)$ and $f_0(980)$ should be mixing
states of $S_8$ and $S_0$ via the mixing angle defined as
\begin{eqnarray}
\left(
  \begin{array}{c}
    f_0(980) \\
    \sigma(600) \\
  \end{array}
\right) & = & \left(
  \begin{array}{cc}
    \cos\theta_S & \sin\theta_S \\
    -\sin\theta_S & \cos\theta_S \\
  \end{array}
\right)\left(
         \begin{array}{c}
           S_8 \\
           S_0 \\
         \end{array}
       \right)
\end{eqnarray}
In terms of the constituents and the mixing angle $\theta_S$, the
physical $\sigma(600)$ and $f_0(980)$ mesons can be expressed as
\begin{eqnarray}
    f_0(980) & = & (\frac{\cos\theta_S}{\sqrt{6}} + \frac{\sin\theta_S}{\sqrt{3}})( | K^+ K^-\rangle + | K^0\bar{K}^0
\rangle ) + (\frac{\sin\theta_S}{\sqrt{3}} -
\frac{2 \cos\theta_S}{\sqrt{6}}) | \eta\eta \rangle \nonumber\\
    \sigma(600) & = & (-\frac{\sin\theta_S}{\sqrt{6}} + \frac{\cos\theta_S}{\sqrt{3}})( | K^+ K^-\rangle + | K^0\bar{K}^0
\rangle ) + (\frac{\cos\theta_S}{\sqrt{3}} + \frac{2
\sin\theta_S}{\sqrt{6}}) | \eta\eta \rangle
\end{eqnarray}
In the case of ideal mixing with $\cos\theta_S = \frac{1}{\sqrt{3}}$
and $\sin\theta_S = \sqrt{\frac{2}{3}}$ one has
\begin{eqnarray}
    f_0(980) & = & \frac{1}{\sqrt{2}} ( | K^+ K^-\rangle + | K^0\bar{K}^0
\rangle ) ; \;\;\;\;\;\;  \sigma(600) =  | \eta\eta \rangle
\end{eqnarray}
This is the case in which the hadronic molecular picture of $f(980)$
was studied in some literature, for example
Refs.~\cite{Branz:2007xp,Branz:2008qm,Branz:2008ha,Branz:2008cb}. In
our present work, we will take the mixing angle as a parameter and
fit it from the data in the following.

In summary, with the above discussions, one can write the scalar
meson nonet in a matrix form as
\begin{eqnarray}
S & = & \frac{1}{\sqrt{2}}\left(
                            \begin{array}{ccc}
                              a_0^0(980) + \frac{1}{\sqrt{6}}S_8 + \frac{1}{\sqrt{3}} S_0 & \sqrt{2} a_0^+(980) & \sqrt{2}K_0^{\ast +}(800) \\
                              \sqrt{2} a_0^-(980) & - a_0^0(980) + \frac{1}{\sqrt{6}}S_8 + \frac{1}{\sqrt{3}} S_0 & \sqrt{2}K_0^{\ast 0}(800) \\
                              \sqrt{2}K_0^{\ast -}(800) & \sqrt{2}\bar{K}_0^{\ast 0}(800) & -\sqrt{\frac{2}{3}}S_8 + \frac{1}{\sqrt{3}} S_0 \\
                            \end{array}
                          \right) \label{matrix:scalar}
\end{eqnarray}
And the ideal mixing reduces to
\begin{eqnarray}
S_{\rm ideal} & = & \frac{1}{\sqrt{2}}\left(
                            \begin{array}{ccc}
                              a_0^0(980) + f_0(980) & \sqrt{2} a_0^+(980) & \sqrt{2}K_0^{\ast +}(800) \\
                              \sqrt{2} a_0^-(980) & - a_0^0(980) + f_0(980) & \sqrt{2}K_0^{\ast 0}(800) \\
                              \sqrt{2}K_0^{\ast -}(800) & \sqrt{2}\bar{K}_0^{\ast 0}(800) & \sqrt{2}\sigma(600) \\
                            \end{array}
                          \right)
\end{eqnarray}

It should be noted that in the hadronic molecular interpretation
there are more quark contents in the relevant scalars compared to
the tetraquark interpretation as shown in Fig.~\ref{fig:nonet}. This
makes it seems that it is difficult to understand the scalar meson
spectrum in the molecular picture. However, at present, we do not
attempt to calculate the scalar meson spectrum but take the scalar
meson masses as input to determine the coupling constants.

The effective Lagrangian, without including the distribution of the
constituents among the molecules, describing the interaction between
the scalar mesons and their constituents can be written as
\begin{eqnarray}
{\cal L}_{S}^{\rm LC} & = & g_{_{S}}^{\rm LC}P^{\dag}(x)S(x)P(x)
\label{int:s:l}
\end{eqnarray}
where the upper index "LC" denotes the local interaction vertex, $S$
denotes the scalar meson matrix which was give in
Eq.~(\ref{matrix:scalar}) and $P$ denotes the constituent
pseudoscalar meson matrix
\begin{eqnarray}
P & = & \left(
          \begin{array}{c}
            K^+ \\
            K^0 \\
            \eta \\
          \end{array}
        \right)
\end{eqnarray}

To include the distribution function which illustrates the
distribution of the constituents in the molecules, one should modify
the interaction Lagrangian ($\ref{int:s:l}$) as
\begin{eqnarray}
{\cal L}_{S}^{\rm NL} & = & g_{_{S}}^{\rm NL}\int d^4y \Phi_{S}
(y^2)P_i^{\dag}(x + \omega_j y)S_{ij}(x)P_j(x - \omega_i y)
\label{int:s:c}
\end{eqnarray}
where the upper index NL denotes the nonlocal interaction vertex.
It should be noted that to keep the $U(1)_{\rm em}$ gauge
invariance, a Wilson's line connecting the charged particles at
different positions should be introduced. The kinematic parameter
$\omega_i$ is defined as
\begin{eqnarray}
\omega_i & = & \frac{M_i}{M_1 + M_2}
\end{eqnarray}
with $M_i$ as the mass of $i-$th constituent. And correlation
function $\Phi(y^2)$ describes the distribution of the constituents
in the molecule. The Fourier transform of the correlation function
reads
\begin{eqnarray}
\Phi(y^2) & = &
\int\frac{d^4p}{(2\pi)^4}\tilde{\Phi}(p^2)e^{-ip\cdot y}
\end{eqnarray}
In the following numerical calculations, an explicit form of
$\tilde{\Phi}(p^2)$ is necessary. Throughout this paper, we will
take the Gaussian form
\begin{eqnarray}
\tilde{\Phi}(p^2) & = & \exp(p^2/\Lambda_S^2) \label{gaussian}
\end{eqnarray}
where the parameter $\Lambda_S$ is a size parameter which
parametrizes the distribution of the constituents inside the
molecule. The magnitude of $\Lambda_S$ is determined by requiring
that the effective coupling constant $g_{_{S}}^{\rm NL}$ should be
stable against it. It should be noted that the choice
(\ref{gaussian}) is not unique. In principle any choice of
$\tilde{\Phi}(p^2)$, as long as it renders the integral convergent
sufficiently fast in the ultraviolet region, is reasonable. In this
sense, $\tilde{\Phi}(p^2)$ can be regarded as a regulator which
makes the ultraviolet divergent integral well defined. In addition,
we would like to point out that, in the limit $\Phi(y^2) \to
\delta^4(y)$, the interaction between the scalar meson and its
constituents becomes a local one, i.e., Eq.~(\ref{int:s:c})
approaches to Eq.~(\ref{int:s:l}).

\subsection{The Effective Coupling Constant Between the Scalar Meson and its Constituents}

\begin{figure}[htbp]
\begin{center}
\includegraphics[scale=0.4]{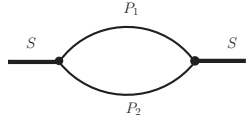}
\end{center}
\caption[Mass operator of scalar meson $S$.]{%
Mass operator of scalar meson $S$. } \label{fig:mass}
\end{figure}

After the discussion on the interaction Lagrangian of the scalar
meson and its constituents, we are ready to calculate the effective
coupling constant $g_{_{S}}$ with the help of the compositeness
condition
\begin{eqnarray}
Z_S & = & 1 - g_{_{S}}^2\Sigma_{S}^\prime (m_S^2;M_1^2,M_2^2) = 1 -
g_{_{S}}^2\frac{d}{d q^2}\Sigma_{S}(q^2;M_1^2,M_2^2)\bigg|_{q^2 =
m_{S}^2} = 0 \label{composite}
\end{eqnarray}
where $M_1$ and $M_2$ are the masses of constituents $P_1$ and
$P_2$, respectively. $g_{_{S}}^2\Sigma_{S}$ is the mass operator of
scalar meson which is depicted in Fig.~\ref{fig:mass}. This
compositeness condition means, after renormalization, the scalar
meson degree of freedoms are removed from the original (bare)
Lagrangian and their dynamics are substituted by the relevant
constituents. In addition, it also indicates, in this model, the
scalar meson cannot arise as a final or initial state since,
according to the LSZ reduction rule, each such external state
contributes a factor $Z_S^{1/2}$ to the physical matrix element so
this kind of matrix elements vanish.

Using the compositeness condition (\ref{composite}) and the
constituents contents of the scalar mesons, one can get the
expressions of the coupling constants as
\begin{eqnarray}
\frac{1}{g^2_{_{a_0^+}}} & = & \Sigma_{a_0^+}^\prime
(m_{a_0}^2;m_{K^+}^2,m_{K^0}^2) \nonumber\\
\frac{1}{g^2_{_{a_0^0}}} & = & \frac{1}{2}[\Sigma_{a_0^0}^\prime
(m_{a_0}^2;m_{K^+}^2,m_{K^+}^2) + \Sigma_{a_0^0}^\prime
(m_{a_0}^2;m_{K^0}^2,m_{K^0}^2)] \nonumber\\
\frac{1}{g^2_{_{K_0^{\ast +}}}} & = & \Sigma_{K_0^{\ast +}}^\prime
(m_{K_0^{\ast}}^2;m_{K^+}^2,m_\eta^2) \nonumber\\
\frac{1}{g^2_{_{K_0^{\ast 0}}}} & = & \Sigma_{K_0^{\ast 0}}^\prime
(m_{K_0^{\ast}}^2;m_{K^0}^2,m_\eta^2) \nonumber\\
\frac{1}{g^2_{_{f_0}}} & = & (\frac{\cos\theta_S}{\sqrt{6}} +
\frac{\sin\theta_S}{\sqrt{3}})^2[\Sigma_{f_0}^\prime
(m_{f_0}^2;m_{K^+}^2,m_{K^+}^2) + \Sigma_{f_0}^\prime
(m_{f_0}^2;m_{K^0}^2,m_{K^0}^2)] \nonumber\\
& & + (\frac{\sin\theta_S}{\sqrt{3}} -
\frac{2\cos\theta_S}{\sqrt{6}})^2\Sigma_{f_0}^\prime
(m_{f_0}^2;m_\eta^2,m_\eta^2) \nonumber\\
\frac{1}{g^2_{_{\sigma}}} & = & (-\frac{\sin\theta_S}{\sqrt{6}} +
\frac{\cos\theta_S}{\sqrt{3}})^2[\Sigma_{\sigma}^\prime
(m_{\sigma}^2;m_{K^+}^2,m_{K^+}^2) + \Sigma_{\sigma}^\prime
(m_{\sigma}^2;m_{K^0}^2,m_{K^0}^2)] \nonumber\\
& & + (\frac{\cos\theta_S}{\sqrt{3}} +
\frac{2\sin\theta_S}{\sqrt{6}})^2\Sigma_{\sigma}^\prime
(m_{\sigma}^2;m_\eta^2,m_\eta^2)
\end{eqnarray}

The mass operator $\Sigma_{S}^\prime(m_S^2;M_1^2,M_2^2)$ can be
calculated by evaluating the standard loop integral. From the
diagram shown in Fig.~\ref{fig:mass}, in the local interaction case
we have
\begin{eqnarray}
\Sigma_{S}^{{\rm LC}~\prime}(m_S^2;M_1^2,M_2^2) & = &
\frac{1}{16\pi^2 m_S^2}
\Big\{\frac{M_1^2-M_2^2}{m_S^2}\ln\frac{M_1}{M_2} - 1 \nonumber\\
& & \;\;\;\;\;\;\;\;\;\;\;\;\;\;\;\;\; + \frac{m_S^2(M_1^2 + M_2^2)
- (M_1^2 -
M_2^2)^2}{m_S^2\sqrt{-\lambda}}\sum_{\pm}\arctan\frac{z_\pm}{\sqrt{-\lambda}}\Big\}
\end{eqnarray}
where
\begin{eqnarray}
z_\pm & = & m_S^2 \pm (M_1^2 - M_2^2) \nonumber\\
\lambda & \doteq & \lambda (m_S^2,M_1^2,M_2^2)
\end{eqnarray}
with $\lambda$ as the K$\ddot{a}$llen function. And similarly, for
the nonlocal interaction case, the standard evaluation leads to
\begin{eqnarray}
\Sigma_{S}^{{\rm NL}~\prime}(m_S^2;M_1^2,M_2^2) & = &
\frac{1}{16\pi^2}\int_0^\infty\int_0^\infty d\alpha d\beta
\frac{P_S}{(1 + \alpha +
\beta)^3}[\frac{d}{dz_m}\tilde{\Phi}^2(z_m)] \label{massderi}
\end{eqnarray}
where
\begin{eqnarray}
P_S & = & \omega_1^2 \alpha + \omega_2^2 \beta + \alpha \beta \nonumber\\
z_m & = & \frac{P_S}{1 + \alpha + \beta}m_S^2 - \alpha M_1^2 - \beta
M_2^2 \label{define:zm}
\end{eqnarray}
It should be noted that the expression (\ref{massderi}) is
independent of the explicit form of $\tilde{\Phi}$. And in the
derivation of (\ref{massderi}) we have applied the Laplace Transform
\begin{eqnarray}
F(s)& = &\int _0^\infty f(s) e^{-sx}dx; ~~~~~~ {\rm for~Re}s > 0
\end{eqnarray}

Up to now, the numerical results of the coupling constants
$g_{a_0^+},g_{a_0^0},g_{K_0^{\ast +}}$ and $g_{K_0^{\ast 0}}$ can be
yielded by inputting the physical masses of the relevant mesons. But
the coupling constants $g_{f_0}$ and $g_{\sigma}$ can not be got
without determining the mixing angle. In the following, we will not
discuss the coupling constant $g_{\sigma}$ because of the large
uncertainty of the sigma meson mass which should be used in the
compositeness condition to determine the coupling constant
$g_\sigma$. For other scalar mesons, the explicitly input masses
are~\cite{Amsler:2008zz} (we adopt the central values of the scalar
meson masses)
\begin{eqnarray}
m_{a_0} & = & 985.1~{\rm MeV};~~~~~~~~~ M_{K_0^{\ast}} =  672~{\rm MeV}; ~~~~~~~~~~ m_{f_0} = 980~{\rm MeV}\nonumber\\
m_{K^+} & = & 493.677~{\rm MeV};~~~~~~ m_{K^0} = 497.648~{\rm
MeV};~~~~~~ m_{\eta} = 547.51~{\rm MeV} \label{mesonmass}
\end{eqnarray}
With these numerical values, we get the following numerical results
for the local interaction coupling constants
\begin{eqnarray}
g_{_{a_0^+}}^{\rm LC} & = & 3.554~{\rm GeV}\;\; ;
\;\;\;\;g_{_{a_0^0}}^{\rm LC} = 3.169~{\rm GeV}\;\; ; \;\;\;\;
g_{_{K_0^{\ast+}}}^{\rm LC} = 12.99~{\rm GeV}\;\; ; \;\;\;\;
g_{_{K_0^{\ast0}}}^{\rm LC} =
13.06~{\rm GeV}\nonumber\\
\frac{1}{g^{\rm LC~2}_{_{f_0}}} & = & 0.1194 \times
(-\frac{\sin\theta_S}{\sqrt{6}} + \frac{\cos\theta_S}{\sqrt{3}})^2 +
0.0117 \times (\frac{\cos\theta_S}{\sqrt{3}} +
\frac{2\sin\theta_S}{\sqrt{6}})^2
\end{eqnarray}

\begin{figure}[htbp]
\begin{center}
\includegraphics[scale=0.4]{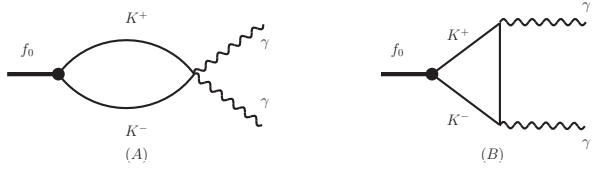}
\end{center}
\caption[Diagrams contribute to the decays of $f_0 \to 2\gamma$.]{%
Diagrams contribute to the decays of $f_0 \to 2\gamma$. }
\label{fig:2gamma}
\end{figure}

In the following, we will fix the coupling constant $ g_{f_0} $ and
mixing angle $ \theta _S $ using the two-photon decay of $f_0$. The
two-photon decay of $ f_0 \to 2\gamma $ has been studied in the
hadronic molecular model in
Refs.~\cite{Giacosa:2007bs,Branz:2007xp,Branz:2008ha}. In the local
interaction case, one should consider the diagrams shown in
Fig.~\ref{fig:2gamma}. And, even in the nonlocal interaction case,
only including these diagrams is enough since, to our experiences,
the contributions from diagrams with photon emerges from the
nonlocal vertex are negligible.

Generally, the width for decay $f_0 \to 2\gamma$ can be expressed as
\begin{eqnarray}
\Gamma(S \to 2\gamma) & = & 16\pi\alpha_{\rm em}^2
\tilde{g}_{_{f_0}}^2 G^2(m_S^2)m_S^3;
\end{eqnarray}
And, because there are only charged kaon mesons in the loop,
$\tilde{g}_{_{f_0}}$ relates to the coupling constant $g_{_{f_0}}$
via relations
\begin{eqnarray}
\tilde{g}_{_{f_0}} & = & g_{_{f_0}}(\frac{\cos\theta}{\sqrt{6}} +
\frac{\sin\theta}{\sqrt{3}})
\end{eqnarray}
For the local interaction case, after standard calculation we get
\begin{eqnarray}
G^{\rm LC}(m_{f_0}^2) & = & \frac{1}{16\pi^2}\int_0^1 dx_1
\int_0^{1-x_1}dx_2 \frac{x_1x_2}{-x_1 x_2 m_{f_0}^2 + m_K^2}
\label{effec:electro:lc}
\end{eqnarray}
And for the nonlocal interaction, we get
\begin{eqnarray}
G^{\rm NL}(m_{f_0}^2) & = &
\frac{1}{16\pi^2}\frac{1}{\Lambda_S^2}\int_0^\infty d\alpha_1
d\alpha_2 d\alpha_3 \frac{1}{(1+\sum_{i=1}^3\alpha_i)^4}(\frac{1}{2}
+ \alpha_1)(\frac{1}{2} + \alpha_2)\Phi(z_R)
\label{effec:electro:nl}
\end{eqnarray}
with
\begin{eqnarray}
z_R & = & -\frac{1}{1+\sum_{i=1}^3\alpha_i}(\frac{1}{2} + \alpha_2 +
\alpha_3)(\frac{1}{2} + \alpha_2)m_{f_0}^2 + (\frac{1}{4} +
\alpha_2)m_{f_0}^2 - \sum_{i=1}^3\alpha_i m_K^2
\end{eqnarray}

Using the average data quoted by PDG~\cite{Amsler:2008zz}
\begin{eqnarray}
\Gamma (f_0 \to 2\gamma) & = & 0.29  ^{+0.07}_{-0.09} {~\rm KeV}
\end{eqnarray}
and the expression (\ref{effec:electro:lc}), we have the numerical
results of the coupling constant $\tilde{g}_{f_0}$ as
\begin{eqnarray}
\tilde{g}_{f_0}^{\rm LC} & = & 2.878~{\rm GeV};\nonumber\\
\tilde{g}_{f_0}^{\rm NL} & = & 2.941 - 2.915~{\rm GeV}; \;\;\; {\rm
for} \;\; \Lambda_S = 3.0 - 4.0 {~\rm GeV}
\end{eqnarray}
which leads to
\begin{eqnarray}
g_{f_0}^{\rm LC} & = & 4.175 {~\rm GeV}; ~~~~~~{\rm for~~} \theta_S = 42^\circ \nonumber\\
g^{\rm LC}_{f_0} & = & 4.182 {~\rm GeV}; ~~~~~~{\rm for~~} \theta_S
= 68^\circ
\end{eqnarray}
In principle, with the help of the decay of $\sigma \to 2\gamma$ one
can determine the exact value of $\theta_S$ if this model can
explain all the data well. But, as we mentioned above, we will not
study the physics of $\sigma$ meson because of its large mass
uncertainty, so that at present we will adopt $\theta_S = 68^\circ$.
We would like to point out that our final results do not depend on
the choice of $\theta_S$ closely, since the two angles lead to the
same effective $f_0KK$ coupling constant and the $\eta\eta$
component plays negligible roles in the quantities we calculated.

Using the same method, in the nonlocal interaction case, one can
also determine the effective coupling constants $g_{_{S}}$. In this
case, since the coupling constants are functions of $\Lambda_S$, one
should determine the magnitude of $\Lambda_S$ at first. Our
principle is that the effective coupling constants should be stable
against $\Lambda_S$. With this criterion in mind, and running
$\Lambda_S$ from $1.0~$GeV to $10.0~$GeV, we find the physical
region of $\Lambda_S$ should be $3.0 - 4.0~$GeV. We list the
numerical results of $g_{_S}$ in Table.\ref{table:coupling} and
compare them with that from the local interaction case and other
literature.

\begin{table}
\caption{\label{table:coupling} Numerical results of the coupling
constants $g_{_S}$ (in unit of GeV). The range of our results is due
to the variation of $\Lambda_S$ from $3.0 - 4.0~$GeV.}

\begin{tabular}{llllllll}
\hline \hline
&  \hspace*{.6cm} $g_{a_0^\pm}$ & \hspace*{.6cm} $g_{a_0^0}$ & \hspace*{.6cm} $g_{K_0^{\ast\pm}}$  & \hspace*{.6cm} $g_{K_0^{\ast0}}$  & \hspace*{.1cm} $g_{f_0} $  & \hspace*{.1cm} $\theta_S $ \\
\hline
 Our results (NL) & \, $ 3.591 \sim 3.577  $  & \, $ 3.252 - 3.227  $  & \, $ 14.13 - 13.73  $  & \, $ 13.23 - 13.81 $  & \, $  4.230 \sim 4.210 $ \,& \, $ 42^\circ $  \,\, \\
~ & \, ~  & \, ~   & \, ~  & \, ~  & \, $ 4.237 - 4.217 $ \,& \, $ 68^\circ $  \,\, \\
\hline
 Our results (LC) & \, $ 3.554  $  & \, $ 3.169  $  & \, $ 12.99 $  &\, $ 13.06 $   & \, $ 4.175 $  & \, $ 42^\circ $  \,\, \\
 ~ & \, ~  & \, ~  & \, ~  &\, ~   & \, $ 4.182 $  & \, $ 68^\circ $  \,\, \\
\hline
 Ref.~\cite{Branz:2007xp} & \, $ ~ $  & \, $ ~ $ & \, $~$   & \, $~$   & \, $ 3.09 $   & \,\,\,\, $ ~ $  \,\, \\
\hline
 Ref.~\cite{Branz:2008ha} & \, $ 3.61 $~(NL)  & \, $ 3.61 $~(NL) & \, $~$   & \, $~$   & \, $ 3.06 $~(NL) & \,\,\,\, $ ~ $  \,\, \\
 ~ & \, $ 3.45 $~(LC)  & \, $ 3.45 $~(LC) & \, $~$   & \, $~$   & \, $ 2.87 $~(LC)   & \,\,\,\, $ ~ $  \,\, \\
\hline Ref.~\cite{Napsuciale:2004au} & \, $ ~ $  & \, $  ~ $ & \, $~$   & \, $~$   & \, $ 3.27 \pm 0.99 $ & \,\,\,\, $ ~ $  \,\, \\
\hline \hline
\end{tabular}
\end{table}

In this table we find in the physical region of $\Lambda_S$ all the
coupling constants calculated in the nonlocal interaction case agree
with that yielded in the local interaction case. However, compared
with Ref.~\cite{Branz:2007xp}, one find that our result for
$g_{_{f_0}}$ is much larger. This is because, in the present
construction, in addition to the $KK$ constituents, $f_0$ also
consists of $\eta\eta$ component so there is the mixing angle
$\theta_S$. When the mixing angle $\theta_S$ is included our
effective coupling constant $g_{_{f_0KK}}$ agrees with that given in
the literature. This is one of the typical properties of the present
work. Concerning the typical property that numerical values of the
coupling constants yielded from the nonlocal interaction case agree
with that from the local interaction case, in the following
calculation, we will adopt the local interaction vertex.

At last we would like to mention that, because of the $K\bar{K}$
components in $f_0$ and $a_0^0$, there is $a_0^0-f_0$ mixing in the
hadronic molecular model. In the present work we will not study the
mixing effect of these two mesons. For persons who interest in this
topic please see Refs.~\cite{Branz:2008ha,Wu:2007jh,Wu:2008hx} and
the relevant references therein.

\subsection{The Leptonic Decays of the Scalar Mesons}

\begin{figure}[htbp]
\begin{center}
\includegraphics[scale=0.4]{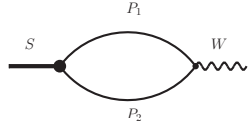}
\end{center}
\caption[Diagram relates to the leptonic decay constant of the charged scalar meson.]{%
Diagram relates to the leptonic decay constant of the charged scalar
meson. } \label{fig:leptondecay}
\end{figure}

Next, we will calculate the scalar meson leptonic decay widths in
the molecular picture. It is well known that, for a pseudoscalar
meson, for example $\pi^+$, the leptonic decay constant $f_\pi$ is
defined by
\begin{eqnarray}
\langle | \bar{u}\gamma_\mu \gamma_5 d| \pi^- (p_\pi)\rangle & = &
-if_\pi p_\pi^\mu
\end{eqnarray}
And since leptons are free from strong interaction, one can express
the width of the pion weak decay $\pi^- \to l \bar{\nu}_l$ in terms
of the leptonic decay constant $f_\pi$ using naive factorization.
Comparing with the data, one can extract the numerical value of
$f_\pi$. On the other hand, if the wave function of pion is known,
we can also calculate this quantity by the standard loop integral.
For our present problem, the scalar mesons are interpreted as
hadronic molecules and the coupling constant between the scalar
meson and its constituents is determined by the compositeness
condition. So that we can calculate the leptonic decay constants of
scalar mesons via the standard loop integral. At the meson level, we
define the leptonic decay constant of the scalar meson via
\begin{eqnarray}
\langle S(p) | P_1 \partial^{^{^{\hspace{-0.2cm} \leftrightarrow}}}
_\mu P_2| 0 \rangle & = & i p_\mu f_s \label{def:lep}
\end{eqnarray}
where $p_\mu$ is the momentum of scalar meson and $P_1$ and $P_2$
are the two constituents of scalar meson $S$. To calculated this
quantity, one should concern the the Feynman diagram depicted in
Fig.~\ref{fig:leptondecay}.

The coupling constants between the weak gauge bosons and
pseudoscalar mesons can be yielded by gauging the nonlinear sigma
model and relating the flavor symmetry of the nonlinear sigma model
to the flavor symmetry of QCD. The relevant terms
are~\cite{HerreraSiklody:1996pm}
\begin{eqnarray}
{\cal L}_{\rm nl\sigma}^{\rm gauged} & = & \frac{F_\pi^2}{4} {\rm
Tr}[D_\mu U (D_\mu U^\dag) + \chi ^\dag U + U^\dag \chi
]\label{nonlinearsigma}
\end{eqnarray}
where $\chi = 2 B M$ with $B$ as a constant relating to the quark
condensation and $M = {\rm diag}(m_u,m_d,m_s)$ as the quark mass
matrix. The field $U(x)$ is expressed in terms of pseudoscalar
fields as
\begin{eqnarray}
U(x) & = & \exp(i\phi(x)/F_\pi) \nonumber\\
\phi(x) & = & \left(
                \begin{array}{ccc}
                  \pi^0 + \frac{1}{\sqrt{3}}\eta  & \sqrt{2}\pi^+ & \sqrt{2}K^+ \\
                  \sqrt{2}\pi^- & -\pi^0 + \frac{1}{\sqrt{3}}\eta & \sqrt{2}K^0 \\
                  \sqrt{2}K^- & \sqrt{2}\bar{K}^0 & - \frac{2}{\sqrt{3}}\eta \\
                \end{array}
              \right)
\end{eqnarray}
where $F_\pi$ is the pion leptonic decay constant with the value
$F_\pi = 92.5{~\rm MeV}$. Under $SU(3)_L\times SU(3)_R$ chiral
transformation meson matrix $U(x)$ transforms as
\begin{eqnarray}
U(x) & \rightarrow & g_R U(x) g_L^\dag
\end{eqnarray}
The covariant derivative is defined as
\begin{eqnarray}
D_\mu U & = & \partial _ \mu U - i A_{R;\mu} U + i U A_{L;\mu}
\end{eqnarray}
$A_{L;\mu}$ and $A_{R;\mu}$ are the gauge fields corresponding to
the gauged left- and right-handed chiral symmetry, respectively.

By matching the chiral symmetry of QCD and the transformation of the
field $U(x)$ one can express these gauge fields in terms of the
electroweak gauge bosons as~\cite{Harada:1995sj}
\begin{eqnarray}
A_{R;\mu} & = & -e Q A_\mu - g \frac{\sin^2\theta_W}{\cos\theta_W}Q Z_\mu \nonumber\\
A_{L;\mu} & = & -e Q A_\mu + g Q_Z Z_\mu +
\frac{g}{\sqrt{2}}(W_\mu^+ Q_W + W_\mu^- Q_W^\dag) \label{def:gauge}
\end{eqnarray}
where $Q$ is the charge matrix of quarks and in the three flavor
case $Q = {\rm diag}(2/3,-1/3,-1/3)$ and $e = g \sin\theta_W$. The
matrices $Q_W$ and $Q_Z$ are defined as
\begin{eqnarray}
Q_W & = & \left(
            \begin{array}{ccc}
              0 & V_{ud} & V_{us} \\
              0 & 0 & 0 \\
              0 & 0 & 0 \\
            \end{array}
          \right); \;\;\;\;\;\;\; Q_Z = \frac{1}{\cos\theta}\left(
            \begin{array}{ccc}
              1/2 & 0 & 0 \\
              0 & -1/2 & 0 \\
              0 & 0 & -1/2 \\
            \end{array}
          \right) - \frac{\sin^2\theta_W}{\cos\theta_W}Q
\end{eqnarray}
where $V_{ud}$ and $V_{us}$ are the appropriate
Cabibbo-Kobayashi-Maskawa matrix elements. And $g$ is the coupling
constant of the $SU(2)_L$ weak gauge group in the standard model and
at the lowest order perturbation theory it is determined by the
Fermi constant and the W boson mass via the
relation~\cite{Amsler:2008zz}
\begin{eqnarray}
G_F & = & \sqrt{2}\frac{g^2}{8m_W^2} = 1.16637(1) \times 10^{-5}
{\rm GeV}^{-2}
\end{eqnarray}
Explicitly, one can get the following $WPP$ interaction Lagrangian
\begin{eqnarray}
{\cal L}_W^{W} & = & -\frac{ig}{4\sqrt{2}} W^+_\mu \Big[
2V_{ud}K^0\partial^{^{^{\hspace{-0.2cm}\leftrightarrow}}}_\mu K^- -
\sqrt{6}V_{us}\eta\partial^{^{^{\hspace{-0.2cm}\leftrightarrow}}}_\mu
K^-  \Big]
\end{eqnarray}

As an example, let us consider the decay of $ a_0^-(980) \to l
\bar{\nu}_l $. Its matrix element, on the one hand, can be expressed
in terms of the leptonic decay constant $f_{a_0^+}$, and on the
other hand, can be calculated directly from the relevant Feynman
diagram in terms of the coupling constant $g_{a_0^+}$, so that
$f_{a_0^+}$ can be expressed as a function of the coupling constant
$g_{a_0^+}$. In this sense the numerical value of $f_{a_0^+}$ is
calculable in the molecular model.

Consider the following effective Hamiltonion
\begin{eqnarray}
H_{\rm eff} & = & \sqrt{2}i G_F
V_{ud}[K^0\partial^{^{^{\hspace{-0.2cm}\leftrightarrow}}}_\mu
K^-][\bar{l}\gamma_\mu P_L \nu_l]
\end{eqnarray}
on the one hand, the relevant matrix element, with the help of
"naive factorization", can be written as
\begin{eqnarray}
i M(a_0^- \to l \bar{\nu}_l) & = & -\sqrt{2}i G_F V_{ud} \langle l,
\bar{\nu}_l |
[K^0\partial^{^{^{\hspace{-0.2cm}\leftrightarrow}}}_\mu
K^-][\bar{l}\gamma_\mu P_L \nu_l]| a_0^-\rangle\nonumber\\
& = & \sqrt{2}G_F V_{ud} p_{a_0^-}^\mu f_{a_0^-} \bar{u}(p_\mu)
\gamma_\mu P_L v(p_{\nu_\mu}) \label{matrixfa}
\end{eqnarray}
On the other hand, this matrix element can be expressed in terms of
the coupling constant $g_{a_0^+}$ and standard loop integral as
\begin{eqnarray}
i M(a_0^- \to l \bar{\nu}_l) & = & -\sqrt{2}G_FV_{ud}\bar{u}(p_\mu)
\gamma_\mu P_L
v(p_{\nu_\mu})\nonumber\\
& & \times
g_{a_0^+}\int\frac{d^4k}{(2\pi)^4}\frac{1}{k^2-m_{K^0}^2}\frac{1}{(k+p)^2-m_{K^+}^2}(2k+p_{a_0^-})_\mu
\label{matrixga}
\end{eqnarray}
Comparing (\ref{matrixfa}) and (\ref{matrixga}) and after some
calculations, one gets the leptonic decay constant of $a_0$ as
\begin{eqnarray}
f_{a_0^-} & = & g_{a_0^+} I(m_{a_0}^2, m_{K^0}^2,
m_{K^+}^2)\label{leptscalar}
\end{eqnarray}
where
\begin{eqnarray}
I(m_{a_0}^2, m_{K^0}^2, m_{K^+}^2) & = & \int_0^1 dx (1-2x)\ln \Big[
\frac{(1-x)m_{K^0}^2 + x m_{K^+}^2 -x(1-x)m_{a_0}^2}{\mu^2}\Big]
\label{intletlc}
\end{eqnarray}
where $\mu$ is the dimensional parameter introduced in dimensional
regularization. It should be noted that, in this expression, the
term $(1-2x)\ln \mu^2$ vanishes after the Feynman parameter
integral. In this sense the the expression (\ref{intletlc}) is scale
independent. With the Eqs.~(\ref{leptscalar},\ref{intletlc}) we get
the numerical result of the leptonic decay constant $f_{a_0^+}$ for
scalar meson $a_0^+$ as
\begin{eqnarray}
f_{a_0^+} & = & g_{a_0^+} I_L(m_{a_0}^2, m_{K^+}^2, m_{K^0}^2) =
0.1530~{\rm MeV}
\end{eqnarray}
Similarly, one can get the leptonic decay constant for $K_0^{\ast+}$
as
\begin{eqnarray}
f_{K_0^{\ast+}} & = & g_{K_0^{\ast+}} I_L(m_{K_0^{\ast}}^2,
m_{K^+}^2, m_{\eta}^2) = 3.463~{\rm MeV}
\end{eqnarray}
For $K_0^{\ast 0}$ meson, one can also define its leptonic decay
constant as~(\ref{def:lep}) and express $f_{K_0^{\ast 0}}$ as
(\ref{leptscalar}). The numerical calculation yields
\begin{eqnarray}
f_{K_0^{\ast0}} & = & g_{K_0^{\ast0}} I_L(m_{K_0^{\ast}}^2,
m_{K^0}^2, m_{\eta}^2) = 3.208~{\rm MeV}
\end{eqnarray}
From this result one can see that the isospin violating effect is
important for the study of $K_0^{\ast}$ mesons in the molecular
model.

We would like to mention that, in contrast to the quark model where
the leptonic decay constants of neutral mesons are identified with
their charged partners, the leptonic decay constants of neutral
molecular scalar mesons  $f_0$ and $a_0^0$ are zero since, equation
(\ref{intletlc}) vanishes in case of $M_1 = M_2$. Physically, this
is because the weak interaction (and the electromagnetic
interaction) is mediated by vector current of pseudoscalar mesons.

With the help of leptonic decay constants calculated above, we can
calculate the leptonic decay of the charged scalar mesons. In terms
of the leptonic decay constants, one can express the partial width
for decay $a_0^- \to l \bar{\nu}_l$ as
\begin{eqnarray}
\Gamma(a_0^- \to l \bar{\nu}_l) & = &
\frac{G_F^2|V_{ud}|^2}{8\pi}f_{_{a_0^+}}^2m_{a_0}m_l^2
\Big(1-\frac{m_l^2}{m_{a_0}^2}\Big)^2
\end{eqnarray}
Along the same derivation, the leptonic decay width of $K_0^{\ast-}
\to l \bar{\nu}_l$ can be expressed as
\begin{eqnarray}
\Gamma(K_0^{\ast-} \to l \bar{\nu}_l) & = &
\frac{6G_F^2|V_{us}|^2}{32\pi}f_{_{K_0^{\ast}}}^2m_{K_0^{\ast}}m_l^2
\Big(1-\frac{m_l^2}{m_{K_0^{\ast}}^2}\Big)^2
\end{eqnarray}

The numerical results are found to be
\begin{eqnarray}
\Gamma(a^- \to e^- \bar{\nu}_e) & = & 3.091 \times 10^{-20}{\rm ~KeV}~~;~~~~~~ \Gamma(a^- \to \mu^- \bar{\nu}_\mu) = 1.306 \times 10^{-15}{\rm ~KeV}\nonumber\\
\Gamma(K_0^{\ast-} \to e^- \bar{\nu}_e) & = & 1.620 \times
10^{-17}{\rm ~KeV}~~;~~~~~~ \Gamma(K_0^{\ast-} \to \mu^-
\bar{\nu}_\mu) = 6.755 \times 10^{-13}{\rm ~KeV}
\end{eqnarray}
From these numerical results we conclude that, in the hadronic
molecular interpretation, it is difficult to search for the leptonic
decays of the charged scalar mesons in the near future. Or
inversely, if the observed leptonic decay widths of scalar mesons
are much larger than the present results, the hadronic molecular
interpretation is suspectable.

\section{Strong Decays of Light Scalar Mesons}

\label{sec:strong}

\begin{figure}[htbp]
\begin{center}
\includegraphics[scale=0.4]{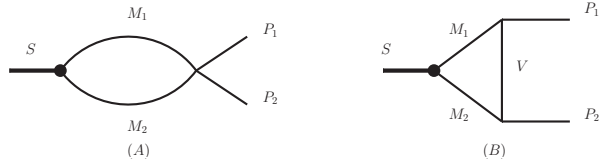}
\end{center}
\caption[Feynman diagrams contributing to the strong decay of $S \to P_1 P_2$ ($M_i$ denotes the constituent meson, $P_i$ denotes the final pseudoscalar meson and $V$ denotes the relevant vector meson).]{%
Feynman diagrams contributing to the strong decay of $S \to P_1 P_2$
($M_i$ denotes the constituent meson, $P_i$ denotes the final
pseudoscalar meson and $V$ denotes the relevant vector meson). }
\label{fig:sppdecay}
\end{figure}

In this section, based on the hadronic molecular explanation, we
will study the strong decays of the light scalar mesons with masses
below 1.0 GeV. Explicitly, we will study the scalar meson to
two-pseudoscalar meson decays, i.e., the decays of $a_0^+ \to \eta
\pi^+, a_0^0 \to \eta \pi^0, K_0^{\ast +} \to K \pi, K_0^{\ast 0}
\to K \pi $ and $f_0 \to \pi \pi$. These processes are important
since they are the dominant channels for the relevant scalar
mesons~\cite{Amsler:2008zz}. To study these decays, one should
consider the two kinds of Feynman diagrams depicted in
Fig.~\ref{fig:sppdecay}. Diagram (A) is from the four-pseudoscalar
meson vertex, while diagram (B) arises from the final state
interaction.

In the explicit calculation, we need the four-pseudoscalar
interaction vertices which give diagram (A) of
Fig.~\ref{fig:sppdecay}. Here we only consider the terms from the
leading order chiral perturbation theory with the explicit chiral
symmetry breaking terms~\cite{Gasser:1983yg,Gasser:1984gg}
\begin{eqnarray}
{\cal L}^{\mathcal{O}(2)} & = & \frac{F_\pi^2}{4} {\rm
Tr}[\partial_\mu U (\partial_\mu U^\dag) + \chi ^\dag U + U^\dag
\chi ]\nonumber\\
& = & \frac{1}{24F_\pi^2} \nonumber\\
& & \times \Big\{ \Big[ 4(\pi^- \partial^{^{^{\hspace{-0.2cm}\leftrightarrow}}}_\mu \bar{K}^0)(K^0 \partial^{^{^{\hspace{-0.2cm}\leftrightarrow}}}_\mu \pi^+) + 4(K^0 \partial^{^{^{\hspace{-0.2cm}\leftrightarrow}}}_\mu \bar{K}^0)(\pi^- \partial^{^{^{\hspace{-0.2cm}\leftrightarrow}}}_\mu \pi^+) + 4(m_K^2 + m_\pi^2) K^0 \bar{K}^0 \pi^+ \pi^- \Big] \nonumber\\
& & \;\;\; + \Big[ 4(\pi^- \partial^{^{^{\hspace{-0.2cm}\leftrightarrow}}}_\mu K^+)(K^- \partial^{^{^{\hspace{-0.2cm}\leftrightarrow}}}_\mu \pi^+) + 4(K^- \partial^{^{^{\hspace{-0.2cm}\leftrightarrow}}}_\mu K^+)(\pi^- \partial^{^{^{\hspace{-0.2cm}\leftrightarrow}}}_\mu \pi^+) + 4(m_K^2 + m_\pi^2) K^+ K^- \pi^+ \pi^- \Big] \nonumber\\
& & \;\;\; + \Big[ 2\sqrt{3} (\eta \partial^{^{^{\hspace{-0.2cm}\leftrightarrow}}}_\mu K^0)(\pi^0 \partial^{^{^{\hspace{-0.2cm}\leftrightarrow}}}_\mu \bar{K}^0) + 2\sqrt{3}(\eta \partial^{^{^{\hspace{-0.2cm}\leftrightarrow}}}_\mu \bar{K}^0)(\pi^0 \partial^{^{^{\hspace{-0.2cm}\leftrightarrow}}}_\mu K^0) + \frac{4}{\sqrt{3}}(m_K^2 - m_\pi^2) K^0 \bar{K}^0 \pi^0 \eta \Big] \nonumber\\
& & \;\;\; + \Big[ 2\sqrt{6} (\eta \partial^{^{^{\hspace{-0.2cm}\leftrightarrow}}}_\mu K^0)(K^- \partial^{^{^{\hspace{-0.2cm}\leftrightarrow}}}_\mu \pi^+) + 2\sqrt{6}(\eta \partial^{^{^{\hspace{-0.2cm}\leftrightarrow}}}_\mu K^-)(K^0 \partial^{^{^{\hspace{-0.2cm}\leftrightarrow}}}_\mu \pi^+) - 4\sqrt{\frac{2}{3}}(m_K^2-m_\pi^2) K^0 K^- \pi^+ \eta  \Big]\nonumber\\
& & \;\;\; + \Big[ 2\sqrt{6} (\eta \partial^{^{^{\hspace{-0.2cm}\leftrightarrow}}}_\mu\bar{K}^0)(K^+ \partial^{^{^{\hspace{-0.2cm}\leftrightarrow}}}_\mu \pi^-) + 2\sqrt{6}(\eta \partial^{^{^{\hspace{-0.2cm}\leftrightarrow}}}_\mu K^+)(\bar{K}^0 \partial^{^{^{\hspace{-0.2cm}\leftrightarrow}}}_\mu \pi^-) - 4\sqrt{\frac{2}{3}}(m_K^2-m_\pi^2) K^+ \bar{K}^0 \pi^- \eta \Big]\nonumber\\
& & \;\;\; + \Big[ 2\sqrt{3} (\eta \partial^{^{^{\hspace{-0.2cm}\leftrightarrow}}}_\mu K^-)(K^+ \partial^{^{^{\hspace{-0.2cm}\leftrightarrow}}}_\mu \pi^0) + 2\sqrt{3} (\eta \partial^{^{^{\hspace{-0.2cm}\leftrightarrow}}}_\mu K^+)(K^- \partial^{^{^{\hspace{-0.2cm}\leftrightarrow}}}_\mu \pi^0) - \frac{4}{\sqrt{3}}(m_K^2 - m_\pi^2) K^+ K^- \pi^0 \eta \Big] \nonumber\\
& & \;\;\; + 4m_\pi^2 \pi^+ \pi^- \eta \eta \Big\} + \cdots
\label{fourpvertex}
\end{eqnarray}
where only terms will be used in our calculation were written down.
And to derive these relations we have used the Gell-Mann-Okubo mass
relation
\begin{eqnarray}
4 m_K^2 & = & 3 m_\eta^2 + m_\pi^2
\end{eqnarray}
The interaction Lagrangian for vector-pseudoscalar-pseudoscalar
meson vertex can be written as
\begin{eqnarray}
{\cal L}_{K^{\ast}K\pi} & = &
\frac{ig_{K^{\ast}K\pi}}{\sqrt{2}}K^{\ast~\dag}_\mu(x)
\vec{\tau}\cdot\vec{\pi}(x)
\partial_{\mu}^{^{^{\hspace{-0.2cm}\leftrightarrow}}}K(x) + {\rm H.c}
\nonumber\\
{\cal L}_{K^{\ast}K\eta} & = &
\frac{ig_{K^{\ast}K\eta}}{\sqrt{2}}K^{\ast~\dag}_\mu(x) \eta(x)
\partial_{\mu}^{^{^{\hspace{-0.2cm}\leftrightarrow}}}K(x) + {\rm H.c}
\end{eqnarray}
where the summation over isospin indices is understood and $A
\partial_{\mu}^{^{^{\hspace{-0.2cm}\leftrightarrow}}} B \equiv A \partial_\mu B - B \partial_\mu
A$. The coupling constant $ g_{K^{\ast}K\pi} $ can be fixed from the
data for the strong decay width $K^{\ast} \to K \pi$. In particular
the strong decay width $\Gamma(K^{\ast} \to K \pi)$ relates to
$g_{K^{\ast}K\pi}$ via
\begin{eqnarray}
\Gamma(K^{\ast} \to K \pi) & = & \frac{g_{K^{\ast}K\pi}^2}{6 \pi
m_{K^{\ast}}^2} P_{\pi K^{\ast}}^3
\end{eqnarray}
where $P_{\pi K^{\ast}}$ is the three-momentum of $\pi$ in the rest
frame of $K^{\ast}$. Using the data for the decay width one can
deduce $g_{K^{\ast}K\pi} =  4.61$~\cite{Amsler:2008zz}. The coupling
constant $g_{K^{\ast}K\eta}$ can be related to the
$g_{K^{\ast}K\pi}$ using the unitary symmetry relation
\begin{eqnarray}
g_{K^{\ast}K\eta} & = &  \frac{F_\pi
\sqrt{3}}{F_\eta}g_{K^{\ast}K\pi} = 6.14
\end{eqnarray}

Generally, the matrix elements corresponding to these two kinds of
diagrams can be written as
\begin{eqnarray}
i M^{(A)}(S \to P_1 P_2) & = & i g_{_{S}} g_{_{M_1M_2P_1P_2}} I^{(A)}(m_S;M_1,M_2; m_1,m_2)\nonumber\\
i M^{(B)}(S \to P_1 P_2) & = & i g_{_{S}}
g_{_{VM_1P_1}}g_{_{VM_2P_2}} I^{(B)}(m_S;m_V;M_1,M_2; m_1,m_2)
\end{eqnarray}
where the notation $m_i$ is the mass of the final state,
$g_{_{M_1M_2P_1P_2}}$ is the coupling constant of four-pseudoscalar
meson vertex which was taken from the leading order Chiral
perturbation theory~\cite{Gasser:1983yg,Gasser:1984gg} and
$g_{_{VM_iP_i}}$ is the coupling constant of
vector-pseudoscalar-pseudoscalar meson vertex. The functions
$I^{(A)}$ and $I^{(B)}$ can be calculated using the standard
technics of loop integral. Concerning the effective Lagrangian
(\ref{fourpvertex}), we will choose $g_{_{M_1M_2P_1P_2}} =
1/(24F_\pi^2)$.

So we formally write the matrix element of $S \to P_1 P_2$ decay as
\begin{eqnarray}
i M(S \to P_1 P_2) & = & i M^{(A)}(S \to P_1 P_2) + i M^{(B)}(S \to
P_1 P_2)
\end{eqnarray}
The decay width, in terms of $I^{(A)}$ and $I^{(B)}$, can be
expressed as
\begin{eqnarray}
\Gamma(S \to P_1 P_2) & = & \frac{g_{_{S}}^2}{8\pi
m_S^2}\Big|\vec{p}_{cm}(m_S^2;m_1^2,m_2^2)\Big|
\nonumber\\
& & \times \Big| g_{_{M_1M_2P_1P_2}} I^{(A)}(m_S;M_1,M_2;
m_1,m_2) \nonumber\\
& & \;\;\;\; + g_{_{VM_1P_1}}g_{_{VM_2P_2}} I^{(B)}(m_S;M_1,M_2;m_V;
m_1,m_2) \Big|^2
\end{eqnarray}
where $\Big|\vec{p}_{cm}(m_S^2;m_1^2,m_2^2)\Big| =
\frac{1}{2m_S}\lambda(m_S^2;m_1^2,m_2^2)$ with $\lambda$ as the
K$\ddot{a}$llen function.

Now, we will calculate the width for strong decay $a_0^+ \to
\pi^+\eta$. At first, let's consider the function $I^{(A)}$. From
the interaction Lagrangian given in Eq.~(\ref{fourpvertex}), one can
see that the four-pseudoscalar-meson vertex consists of two parts:
the part includes derivative and the other part without derivative.
So that we can formally do the decomposition
\begin{eqnarray}
I^{(A)}(m_{a_0};m_{K^+},m_{K^0}; m_\eta,m_{\pi^+}) & = &
g_{P_1P_2P_3P_4}^{D}I^{(A)}_{D}(m_{a_0};m_{K^+},m_{K^0}; m_\eta,m_{\pi^+}) \nonumber\\
& & + g_{P_1P_2P_3P_4}^{ND}I^{(A)}_{ND}(m_{a_0};m_{K^+},m_{K^0};
m_\eta,m_{\pi^+})
\end{eqnarray}
where the upper indices $D$ and $ND$ denote the contributions from
Lagrangian with derivative and without derivative terms,
respectively. After explicit calculation, we get
\begin{eqnarray}
I_D^{(A)}(m_{a_0};m_{K^+},m_{K^0}; m_\eta,m_{\pi^+}) & = & \int\frac{d^4k}{(2\pi)^4}\mathcal{F}(k^2)\frac{i}{k^2-m_K^2}\frac{i}{(k+p)^2-m_{K^0}^2} \\
& &  \;\;\;\;\;\;\;\;\;\;\;\; \times \Big[ [(k+p)+p_1]\cdot(k-p_2)+(p_1-k)\cdot[-p_1-(k+p)] \Big]\nonumber\\
I_{ND}^{(A)}(m_{a_0};m_{K^+},m_{K^0}; m_\eta,m_{\pi^+}) & = &
\int\frac{d^4k}{(2\pi)^4}\mathcal{F}(k^2)\frac{i}{k^2-m_K^2}\frac{i}{(k+p)^2-m_{K^0}^2}
\end{eqnarray}
where $p$ is the momentum of the incoming $a_0^+$ meson, and $p_1$
and $p_2$ are momentum of the outgoing $\eta$ and $\pi^+$,
respectively. Since the momentum integral is divergent, the form
factor ${\cal F}(q^2)$ was introduced to suppress the divergence.
Explicitly, the momentum-dependence of the form factor is
\begin{eqnarray}
{\cal F}(q^2) & = &
\Big(\frac{\Lambda^2-m_K^2}{\Lambda^2-q^2}\Big)^n
\end{eqnarray}
where $n=1, 2$ correspond to the monopole and dipole forms,
respectively~\cite{Machleidt:1987hj}. Through out the following
calculation we will select the dipole form since the above integral
is quadratically divergent, i.e., $n = 2$. For the parameter
$\Lambda$, since we only include the diagrams with the exchanged
mass up to $m_{_{K^{\ast}}}$, we will take the typical value
$\Lambda = 1.0~$GeV. In Appendix.~\ref{App:int}, we express
$I_{D}^{(A)}$ and $I_{ND}^{(A)}$ in terms of the standard n-point
scalar and tensor integrals.

For diagram (B) we have
\begin{eqnarray}
I^{(B)} (m_{a_0};m_{K^+},m_{K^0}; m_\eta,m_{\pi^+}) & = & \int\frac{d^4k}{(2\pi)^4}{\cal F}(k^2)\frac{i}{(k-p_1)^2-m_K^2}\frac{i}{(k+p_2)^2-m_{K^0}^2}\frac{-i}{k^2-m_{K^{\ast}}^2}\nonumber\\
& & \times (2p_1-k)_\mu [g_{\mu\nu} - \frac{k_\mu
k_\nu}{m_{K^\ast}^2}](k+2p_2)_\nu
\end{eqnarray}
which is also expressed in terms of the standard n-point scalar and
tensor integrals in Appendix.~\ref{App:int}.

Substituting the relevant masses and coupling constants and with the
help of the software package LoopTools~\cite{Hahn:1998yk}, we get
the width for the strong decay $a_0^+ \to \pi^+ \eta$ as
\begin{eqnarray}
\Gamma(a_0^+ \to \pi^+ \eta) & = & 58.81{~\rm MeV}
\end{eqnarray}
Along the same method, one can get the following decay width
\begin{eqnarray}
\Gamma(a_0^0 \to \pi^0 \eta) & = & 59.21{~\rm MeV}
\end{eqnarray}
Similarly, for other strong decay widths we have
\begin{eqnarray}
\Gamma(K_0^{\ast +} \to K^0 \pi^+) & = & 11.06{~\rm MeV}\\
\Gamma(K_0^{\ast 0} \to K^- \pi^+) & = & 12.37{~\rm MeV}\\
\Gamma(f_0 \to \pi^+ \pi^-) & = & 30.65{~\rm MeV}
\end{eqnarray}
It should be noted that the analytic forms of diagram (A) in
Fig.~\ref{fig:sppdecay} for these three processes are different from
that of $a_0 \to \eta \pi$ decay we listed in
Appendix.~\ref{App:int}. For simplicity, we will not write it down
explicitly.

With the help of the isospin relation, we get the following decay
widths
\begin{eqnarray}
\Gamma(K_0^{\ast +} \to K \pi) & = & \Gamma(K_0^{\ast +} \to K^+
\pi^0) + \Gamma(K_0^{\ast +} \to K^0 \pi^+) =
\frac{3}{2}\Gamma(K_0^{\ast +} \to K^0 \pi^+) = 16.59{~\rm MeV}  \nonumber\\
\Gamma(K_0^{\ast 0} \to K \pi) & = & \Gamma(K_0^{\ast 0} \to K^+
\pi^-) + \Gamma(K_0^{\ast 0} \to K^0 \pi^0) =
\frac{3}{2}\Gamma(K_0^{\ast 0} \to K^+
\pi^-) = 18.56{~\rm MeV} \nonumber\\
\Gamma(f_0 \to \pi \pi) & = &\Gamma(f_0 \to \pi^+ \pi^-) +
\frac{1}{2}\Gamma(f_0 \to \pi^0 \pi^0) = 45.98{~\rm MeV}
\end{eqnarray}

We summarize our results and compare them with that given in the
literature in Table.~\ref{table:strong}. From this table we see our
results for $\Gamma(f_0 \to \pi \pi)$ and $\Gamma(a_0 \to \eta \pi)$
agree with that of Ref.~\cite{Branz:2008ha}, and the tiny
differences can be understood by concerning that the
scalar-pseudoscalar-pseudoscalar vertices used in the present work
are local one but that in Ref.~\cite{Branz:2008ha} are nonlocal.
This indirectly indicates that the scale $\Lambda = 1.0~$GeV we
chose is reasonable. For the $f_0 \to \pi\pi$ decay, our result is
consistent with that from both the $q\bar{q}$ and tetraquark
interpretations and all the results are consistent with that given
in PDG. For the $a_0 \to \eta \pi$ decays, the results from both the
hadronic molecular interpretation and tetraquark interpretation are
consistent with the data but hadronic interpretation leads to a
smaller result than the tetraquark interpretation. At last, let us
turn to the results of $K_0^{\ast}$ decays. One can see that our
result is much smaller than the data and other theoretical
approaches. In fact, concerning our yielded result is based on the
central value of $K_0^{\ast}$ mass, $m_{K_0^{\ast}} =
672~$MeV~\cite{Amsler:2008zz}, we varied $m_{K_0^{\ast}}$ up to
$1.0~$GeV as a check and it is found that $\Gamma(K_0^{\ast} \to K
\pi) < 70~$MeV which is still much smaller than the data and other
approaches. In this sense, it is difficult to arrange the scalar
mesons with masses below 1.0 GeV into the same nonet.

At last, we would like to mention that, in contrast to the hadronic
interpretation, the the large decay widths of $K_0^\ast$ in
tetraquark picture can be easily understood at the quark level. At
the quark level, in the tetraquark interpretation $K_0^\ast \sim
[qq][\bar{q}\bar{s}]$ (with $q$ as unflavored quark), it can easily
decay into $K\pi$ by interchanging a pair of quark and antiquark and
this process is OZI rule superallowed. However, in the hadronic
molecular model, because of the $s\bar{s}$ component of $\eta$
meson, $K_0^{\ast} \sim K\eta \sim [q\bar{s}][s\bar{s}]$, this
process happens via an annihilation of a strange quark and
antistrange quark, with subsequent $q\bar{q}$ creation, i.e.,
$[q\bar{s}][s\bar{s}] \to [qg\bar{s}] \to [q\bar{q}] + [q\bar{s}]$,
so this process is OZI rule subdominant.

\begin{table}
\caption{\label{table:strong} Numerical results of the widths for
the strong decays $S \to PP $ (in unit of MeV).}

\begin{tabular}{llllllll}
\hline \hline
 &  \hspace*{.1cm}  Our results & \hspace*{.1cm} \cite{Branz:2008ha} & \hspace*{.1cm} \cite{Oller:1998hw} & \hspace*{.1cm} \cite{Scadron:2003yg} & \hspace*{.1cm} \cite{Giacosa:2006rg} & \hspace*{.1cm} PDG~\cite{Amsler:2008zz} \\
\hline
Meson structure &  \hspace*{.1cm} Molecule & \hspace*{.1cm} Molecule & \hspace*{.2cm} hadronic  & \hspace*{.1cm} $q\bar{q}$  & \hspace*{.1cm} $q^2\bar{q}^2 $  & \hspace*{.1cm} ~ \\
\hline
$f_0 \to \pi\pi $ & \, $ 45.98  $  & \, $ 57.4  $  & \, $ 18.2  $  & \, $ 53 $  & \, $ 58 - 136 $ \,& \, $40 - 100 $ \,\, \\
\hline
$a_0^+ \to \pi^+ \eta $ & \, $ 58.81  $  & \, $61.0$   & \, $ 21.0 $  & \, $ 138 $  & \, $ 98 $ \,& \, $50 - 100 $  \,\, \\
\hline
$a_0^0 \to \pi^0 \eta $ & \, $ 59.21  $  & \, $ 61.0 $  & \, $ 21.0 $  &\, $ 138 $   & \, $ 98 $  & \, $50 - 100 $ \,\, \\
\hline
$K_0^{\ast+} \to K \pi $ & \, $ 16.59 $  & \, ~  & \, $ 500.0 $  &\, $ 193 $ & \, $ 241 - 251 $  & \, $550 \pm 34 $  \,\, \\
\hline
$K_0^{\ast0} \to K \pi $ & \, $ 18.56 $  & \, $ ~ $ & \, $ 500.0 $   & \, $ 193 $   & \, $ 241 - 251 $   & \,\,\,\, $550 \pm 34 $ \,\, \\
\hline \hline
\end{tabular}
\end{table}

\section{The productions of neutral scalar mesons $f_0$ and $a_0$ in the radiative decays of $\phi$ meson}

\label{sec:radiativephi}

\begin{figure}[htbp]
\begin{center}
\includegraphics[scale=0.45]{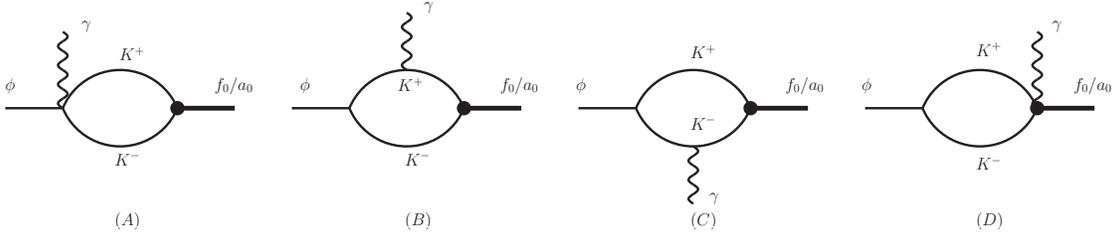}
\end{center}
\caption[Diagrams contributing to the $\phi \to
f_0/a_0\gamma$ Decay (without intermediate resonance contribution).]{%
Diagrams contributing to the $\phi \to f_0/a_0\gamma$ Decay (without
intermediate resonance contribution). } \label{fig:phigammac}
\end{figure}

\begin{figure}[htbp]
\begin{center}
\includegraphics[scale=0.4]{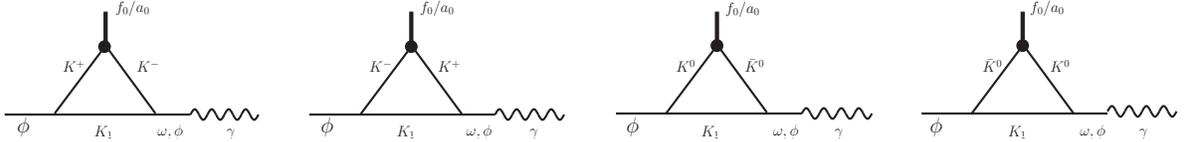}
\end{center}
\caption[VMD diagrams contributing to the $\phi \to
f_0/a_0\gamma$ Decay.]{%
VMD diagrams contributing to the $\phi \to f_0/a_0\gamma$ Decay. }
\label{fig:phigammar}
\end{figure}

\begin{figure}[htbp]
\begin{center}
\includegraphics[scale=0.4]{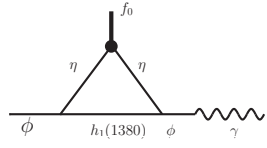}
\end{center}
\caption[VMD diagrams contributing to the $\phi \to
f_0\gamma$ Decay.]{%
VMD diagrams contributing to the $\phi \to f_0\gamma$ Decay
(contribution from the $\eta$ constituent of $f_0$). }
\label{fig:phigammae}
\end{figure}

In this section, we will  study the productions of the scalar mesons
$f_0$ and $a_0$ in the radiative decay of vector meson $\phi$ with
including the intermediate axial-vector mesons and $\eta\eta$
component of $f_0$ meson. These processes are important because they
have long been accepted as a potential route to reveal the natures
of scalar mesons $f_0$ and $a_0$. To study these processes, three
classes of diagrams shown in Figs.~\ref{fig:phigammac},
\ref{fig:phigammar} and \ref{fig:phigammae} should be included. In
Fig.~\ref{fig:phigammac}, diagram $(A)$ arises from the gauge of the
derivative coupling of vector-pseudoscalar-pseudoscalar meson, while
diagrams $(B)$ and $(C)$ are from the gauge of kinetic terms of the
charged constituents. The diagrams in Fig.~\ref{fig:phigammar}
contribute to both $f_0$ and $a_0$ productions from the intermediate
axial-vector mesons and the Fig.~\ref{fig:phigammae} only
contributes to $f_0$ production due to its $\eta\eta$ component. It
should be noted that the diagrams like that in
Fig.~\ref{fig:phigammar} but substituting the axial-vector mesons
with vector mesons will not be considered in the present work since
the sum of the $\phi$ and $\omega$ contributions almost
cancel~\cite{Palomar:2003rb}.

Before explicit calculation, we will discuss the interaction
Lagrangian to be used in the following. At first, the $\phi KK$
interaction Lagrangian can be effectively written as
\begin{eqnarray}
{\cal L}_{\phi KK} & = & ig_{\phi KK}\phi_\mu K^{\dag}(x)
\partial_{\mu}^{^{^{\hspace{-0.2cm}\leftrightarrow}}}K(x)
\end{eqnarray}
Using the data for the strong decay width $\Gamma(\phi \to K^+ K^- =
4.26{~\rm MeV} \times 49.2\% = 2.096{~\rm
MeV}$~\cite{Amsler:2008zz}, one can fix the coupling constant
$g_{\phi KK}$. In particular the strong decay width $\Gamma(\phi \to
K^+ K^- )$ relates to $g_{\phi K K}$ via
\begin{eqnarray}
\Gamma(\phi \to K K) & = & \frac{g_{\phi K K}^2}{6 \pi m_{\phi}^2}
P_{\phi K}^3
\end{eqnarray}
where $P_{\phi K}$ is the three-momentum of $K$ in the rest frame of
$\phi$. Using the masses for the relevant mesons one can deduce
$g_{\phi KK} =  4.48$.

The interaction Lagrangian between photon and vector meson is given
by~\cite{Harada:2003jx}
\begin{eqnarray}
{\cal L}_{V\gamma} & = & - 4 F_\pi^2 e g_{\rho\pi\pi} A_\mu {\rm
Tr}[Q V_\mu] = - \frac{2 F_\pi^2 e g_{\rho\pi\pi}}{3} A_\mu
\omega_\mu + \frac{2\sqrt{2} F_\pi^2 e g_{\rho\pi\pi}}{3} A_\mu
\phi_\mu
\end{eqnarray}
where $Q = (2/3,-1/3,-1/3)$ is the quark electric charge matrix and,
the vector meson matrix is chosen as
\begin{eqnarray}
V_\mu & = &  \frac{1}{2}\left(
                \begin{array}{ccc}
                  \rho^0 + \omega  & \sqrt{2}\rho^+ & \sqrt{2}K^{\ast+} \\
                  \sqrt{2}\rho^- & -\rho^0 + \omega & \sqrt{2}K^{\ast0} \\
                  \sqrt{2}K^{\ast-} & \sqrt{2}\bar{K}^{\ast0} & \sqrt{2}\phi \\
                \end{array}
              \right)_\mu
\end{eqnarray}
and $g_{\rho\pi\pi} = 5.98$ is the universal $\rho\pi\pi$ coupling
constant.

To include the axial-vector mesons, according to
PDG~\cite{Amsler:2008zz}, one should consider two nonets with
quantum numbers $J^{PC} = 1^{+-}$ and $1^{++}$. Explicitly, they can
be written in $U(3)$ matrix forms as
\begin{eqnarray}
A_\mu & = &  \left(
                \begin{array}{ccc}
                  a_1^0 + f_1(1285)  & \sqrt{2}a_1^+ & \sqrt{2}K_{1A}^{+} \\
                  \sqrt{2}a_1^- & - a_1^0 + f_1(1285) & \sqrt{2}K_{1A}^{0} \\
                  \sqrt{2}K_{1A}^{-} & \sqrt{2}\bar{K}_{1A}^{0} & f_1(1420) \\
                \end{array}
              \right)_\mu \nonumber\\
B_\mu & = &  \left(
                \begin{array}{ccc}
                  b_1^0 + h_1(1170)  & \sqrt{2}b_1^+ & \sqrt{2}K_{1B}^{+} \\
                  \sqrt{2}b_1^- & - b_1^0 + h_1(1170) & \sqrt{2}K_{1B}^{0} \\
                  \sqrt{2}K_{1B}^{-} & \sqrt{2}\bar{K}_{1B}^{0} & h_1(1380) \\
                \end{array}
              \right)_\mu
\end{eqnarray}
where $A_\mu$ and $B_\mu$ are the axial-vector matrices with quantum
numbers $J^{PC} = 1^{++}$ and $1^{+-}$, respectively. The mixture of
$K_{iA}$ and $K_{1B}$ with approximate $45^\circ$ mixing angle gives
the physical states $K_1(1270)$ and $K_1(1400)$. Explicitly
\begin{eqnarray}
K_1(1270) & = & \frac{1}{\sqrt{2}}(K_{1B} - i K_{1A} )\nonumber\\
K_1(1400) & = & \frac{1}{\sqrt{2}}(K_{1B} + i K_{1A} )
\end{eqnarray}

Due to the $C$ parity, the interaction Lagrangian for the
axial-vector-vector-pseudoscalar couplings have the following forms
\begin{eqnarray}
{\cal L}_{AVP} & = & i g_{_{AVP}}{\rm Tr}\Big[\; A_\mu [ \; V_\mu, P \; ]\Big]\nonumber\\
{\cal L}_{BVP} & = & g_{_{BVP}}{\rm Tr}\Big[\; B_\mu \{\; V_\mu, P
\; \}\Big]
\end{eqnarray}
where $i$ in front of $g_{_{AVP}}$ was added to keep the Hermitian
of the Lagrangian. Using the concrete matrix forms, one can write
down the interaction vertices that we are interested in
\begin{eqnarray}
{\cal L}_{AVP} 
& = &  \sqrt{2}g_{_{AVP}}  \Big[K_{1;\mu}^+(1270)\rho_\mu^0 K^- - K_{1;\mu}^+(1400)\rho_\mu^0 K^- \nonumber\\
& & \;\;\;\;\;\;\;\;\;\;\;\;\;\;\; + \sqrt{2}K_{1;\mu}^+(1270)\rho_\mu^- \bar{K}^0  - \sqrt{2}K_{1;\mu}^+(1400)\rho_\mu^- \bar{K}^0 \Big] \nonumber\\
& & + \sqrt{2} g_{_{AVP}} \phi_\mu \Big[K_{1;\mu}^+(1400)K^--K_{1;\mu}^+(1270) K^- + K_{1;\mu}^0(1400) \bar{K}^0 - K_{1;\mu}^0(1270) \bar{K}^0 \Big] \nonumber\\
& & - \sqrt{2} g_{_{AVP}} \omega_\mu \Big[K_{1;\mu}^+(1400)K^--K_{1;\mu}^+(1270) K^- + K_{1;\mu}^0(1400) \bar{K}^0 - K_{1;\mu}^0(1270) \bar{K}^0 \Big] + {\rm H.c.}\nonumber\\
{\cal L}_{BVP} 
& = & \sqrt{2}g_{_{BVP}}  \Big[K_{1;\mu}^+(1400)\rho_\mu^0 K^- + K_{1;\mu}^+(1270)\rho_\mu^0 K^-\nonumber\\
& &  \;\;\;\;\;\;\;\;\;\;\;\;\;\;\; + \sqrt{2}K_{1;\mu}^+(1400)\rho_\mu^- \bar{K}^0 + \sqrt{2}K_{1;\mu}^+(1270)\rho_\mu^- \bar{K}^0 \Big] \nonumber\\
& & + \sqrt{2}g_{_{BVP}}\phi_\mu \Big[
K_{1;\mu}^+(1400)K^- + K_{1;\mu}^+(1270)K^-  + K_{1;\mu}^0(1400)\bar{K}^0 + K_{1;\mu}^0(1270) \bar{K}^0 \Big]\nonumber\\
& & + \sqrt{2}g_{_{BVP}}\omega_\mu \Big[
K_{1;\mu}^+(1400)K^- + K_{1;\mu}^+(1270) K^- + K_{1;\mu}^0(1400)\bar{K}^0 + K_{1;\mu}^0(1270) \bar{K}^0 \Big]  + {\rm H.c.}\nonumber\\
& & - \frac{4g_{_{BVP}}}{\sqrt{3}}\phi_\mu h_{1;\mu}(1380) \eta +
\frac{2g_{_{BVP}}}{\sqrt{3}}\omega_\mu  (
b_{1;\mu}^0+h_{1;\mu}(1170) \eta
\end{eqnarray}
The coupling constant $g_{_{K_1VP}}$ can be determined by the decays
of $K_1 \to \rho K$ via the expression
\begin{eqnarray}
\Gamma(K_1 \to VP) & = & \frac{g_{_{K_1VP}}^2}{24\pi
m_{_{K_1}}^2}{\rm P}_{K_1V}\Big[ 3 + \frac{1}{ m_{_{V}}^2}{\rm
P}_{K_1V}^2\Big]
\end{eqnarray}
where $g_{_{K_1VP}}$ is the relevant coupling constant and ${\rm
P}_{K_1V}$ is the three momentum of $V-$meson in the rest frame of
$K_1-$meson. Explicitly, $g_{_{K_1(1270)^+\rho^0 K^-}} =
\sqrt{2}(g_{_{AVP}} + g_{_{BVP}})$ and $g_{_{K_1(1400)^+\rho^0 K^-}}
= \sqrt{2}(g_{_{BVP}} - g_{_{AVP}})$. Using the central values of
the branching ratio and total widths from PDG~\cite{Amsler:2008zz}
$\Gamma(K_1(1270) \to \rho K) = 37.8{~\rm MeV}$ and
$\Gamma(K_1(1400) \to \rho K) = 5.22{~\rm MeV}$ one can get
\begin{eqnarray}
g_{_{K_1(1270)^+\rho^0 K^-}} & = & 3.52~{\rm GeV}; \;\;\;\;
g_{_{K_1(1400)^+\rho^0 K^-}} = 0.54~{\rm GeV}
\end{eqnarray}
which lead to
\begin{eqnarray}
g_{_{AVP}} & = & 1.05~{\rm GeV}; \;\;\;\; g_{_{BVP}} = 1.44~{\rm
GeV}
\end{eqnarray}

We would like to mention that, in the following calculation, we will
not include the diagrams arise from the $\phi-\omega$ mixing. This
is because, the interaction Lagrangian describing the $\phi-\omega$
mixing is~\cite{Urech:1995ry}
\begin{eqnarray}
{\cal L}_{\phi\omega} & = & \Theta \phi_\mu \omega_\mu
\end{eqnarray}
where the coupling constant $\Theta$ is determined by the
relation~\cite{Bramon:1997va}
\begin{eqnarray}
\tilde{\epsilon} & = & \frac{\Theta}{m_\phi^2-m_\omega^2}  = 0.059
\pm 0.004
\end{eqnarray}
all the diagrams with $\phi-\omega$ are suppressed by factor
$\tilde{\epsilon}$ compared with the diagrams without $\phi-\omega$
mixing.

With the above discussions, we can calculate the decay width now.
Generally, due to the gauge invariance, one can write the total
matrix element as
\begin{eqnarray}
iM(\phi(p) \to S(p^\prime) \gamma(q)) & = & i
\epsilon_\mu(p)\epsilon_\nu^{\ast}(q)(g_{\mu\nu}p\cdot q - q_\mu
p_\nu)\; e \; G_{\phi S \gamma}
\end{eqnarray}
and the effective coupling constant $G_{\phi S \gamma}$ consists of
two parts
\begin{eqnarray}
G_{\phi S \gamma} & = & G_{\phi S \gamma}^{c} + G_{\phi S
\gamma}^{r}
\end{eqnarray}
where $G_{\phi S \gamma}^{c}$ is from Fig.~\ref{fig:phigammac}, and
$G_{\phi S \gamma}^{r}$ is from Figs.~\ref{fig:phigammar} and
\ref{fig:phigammae}. The decay width can be expressed as
\begin{eqnarray}
\Gamma(\phi \to S \gamma) & = & \frac{\alpha_{\rm em}}{3}|G_{\phi S
\gamma}|^2 P_\gamma^{\ast~3}
\end{eqnarray}
where $P_\gamma^{\ast} = (m_\phi^2 - m_S^2)/(2 m_\phi)$ is the
three-momentum of the decay products.

With these discussions and selecting $\Lambda = 1.5~$GeV which means
the resonances with masses below $1.5~$GeV were included, one can do
the numerical calculation. Using the effective coupling constants,
we get the decay widths from the contact diagrams and the resonances
as
\begin{eqnarray}
\Gamma(\phi \to f_0 \gamma)^{c} & = & 3.080 \times 10^{-4}{~\rm MeV}\;\; ; \;\; \Gamma(\phi \to a_0 \gamma)^{c} = 2.329 \times 10^{-4}{~\rm MeV}\nonumber\\
\Gamma(\phi \to f_0 \gamma)^{r} & = &  6.670 \times 10^{-13}{~\rm
MeV}\;\; ; \;\; \Gamma(\phi \to a_0 \gamma)^{r} = 3.353 \times
10^{-13}{~\rm MeV}
\end{eqnarray}
From these results we see, compared with the contact diagrams, the
contribution from the axial-vector resonances is negligible. In
summary we have the total decay widths and the branching ratios
(using the total width $\Gamma_\phi =
4.46$~MeV~\cite{Amsler:2008zz}) as
\begin{eqnarray}
\Gamma(\phi \to f_0 \gamma) & = & 3.081 \times 10^{-4}{~\rm MeV};
\;\;\;\;\;\; \Gamma(\phi \to a_0 \gamma) = 2.329 \times 10^{-4}{~\rm
MeV} \nonumber\\
{\rm Br}(\phi \to f_0 \gamma) & = & 6.907 \times 10^{-5};
\;\;\;\;\;\;\;\;\;\;\;\;\;\; {\rm Br}(\phi \to a_0 \gamma) = 5.223
\times 10^{-5}
\end{eqnarray}
and the ratio
\begin{eqnarray}
R & = & \frac{\Gamma(\phi \to f_0 \gamma)}{\Gamma(\phi \to a_0
\gamma)} = 1.32
\end{eqnarray}

\begin{table}

\caption{\label{table:brphisg}  The branch ratio of the $\phi \to
f_0/a_0 \gamma$ decays. }

\begin{tabular}{lllll}
\hline \hline \hspace*{.1cm} Decay modes \hspace*{.2cm}
& \hspace*{.1cm} Present results \hspace*{.1cm}& \hspace*{.2cm} SND\hspace*{.1cm} & \hspace*{.2cm} CLOE\hspace*{.1cm} & \hspace*{.2cm} CMD-2 \hspace*{.2cm} \\
\hline \,\, ${\rm Br}(\phi \to f_0 \gamma) $ & \,\,\, $ 6.91 \times 10^{-5} $ \,\,\,& \,\,\,\, $3.5 \times 10^{-4}$~\cite{Achasov:2000ym} \,\,\,& \,\,\,\, $(4.47 \pm 0.21)\times 10^{-4}$\cite{Aloisio:2002bt}\,\,\,& \,\,\,\, $2.90 \times 10^{-4}$\cite{Akhmetshin:1999di}\,\,\, \\
\,\, ${\rm Br}(\phi \to a_0 \gamma) $ & \,\,\, $ 5.22 \times 10^{-5}$\,\,\,& \,\,\,\, $8.8 \times 10^{-5}$~\cite{Achasov:2000ku} \,\,\,& \,\,\,\, $(7.4 \pm 0.7)\times 10^{-5}$ \cite{Aloisio:2002bsa}\,\,\,& \,\,\,\, $9.0\times 10^{-5}$ \cite{Akhmetshin:1999di}\,\,\, \\
\hline \,\, $\frac{{\rm Br}(\phi \to f_0 \gamma)}{{\rm Br}(\phi \to
a_0 \gamma)} $ & \,\,\, $ 1.32 $ \,\,\,& \,\,\,\, $ 3.98 $ \,\,\,& \,\,\,\, $ 6.04 $ \,\,\,& \,\,\,\, $ 3.22 $ \,\,\, \\
\hline \hline
\end{tabular}
\end{table}

In Table.~\ref{table:brphisg}, we compare our results with that from
the data. From this table one can see that, for the decay $\phi \to
a_0 \gamma$, the present branching ratio is consistent with the
observed data while, that for the decay $\phi \to f_0 \gamma$ is
much smaller than data.  This fact makes $R = {\rm Br}(\phi \to f_0
\gamma)/{\rm Br}(\phi \to a_0 \gamma) = 1.32$ disagree with the
data. Actually, as mentioned in Ref.~\cite{Close:2002zu}, this is
one of the main problems in the hadronic interpretation of the light
scalar mesons. Another observation is that, the final state
interaction gives a negligible contribution to these decays which
agrees with the conclusion given in Ref.~\cite{Palomar:2003rb}.

\section{Discussions and Conclusions}

\label{sec:con}

In this paper, in the framework of effective Lagrangian approach, we
studied the properties of the light scalar mesons with masses below
1.0 GeV in the hadronic molecular interpretation.

To determine the coupling constant $g_{_S}$ between the scalar
molecule and its constituents we applied the compositeness condition
which has been used in our previous works. We found the numerical
results of $g_{_S}$ are consistent with that given in the
literature. With the yielded coupling constant $g_{_S}$ we
calculated the leptonic decay constants and leptonic decay widths of
scalar mesons. Our numerical results show that the leptonic decays
of scalar mesons are not observable in the near future experiments.

The calculations of the strong decays conclude that the decay widths
for $f_0 \to \pi\pi$ and $a_0 \to \eta \pi$ is consistent with the
observed data while width for $K_0^{\ast} \to K \pi$ decay is much
smaller than the data, even the ambiguity from the mass is
considered. This observation shows that the hadronic molecular
interpretation of $K_0^{\ast}$ is disputable and, concerning the
$f_0 \to \pi\pi$ and $a_0 \to \eta \pi$ decays are consistent with
the data, the classification of the scalar mesons with masses below
$1.0~$GeV to form a nonet is disputable.

To study the productions of $f_0$ and $a_0$ in the radiative decays
of $\phi$ meson, we included the contributions from the final state
interaction, i.e., the contributions from the intermediate
axial-vector mesons. The explicit calculation shows that the
dominant contribution is from the contact diagrams of kaon loops and
axial-vector mesons play a negligible role in these processes. The
branching ratio for $\phi \to a_0 \gamma $ is consistent with the
data while that for $\phi \to f_0 \gamma $ is much smaller than the
data. This is another problem of the hadronic molecular model.

We would like to say that, if the isosinglet pseudoscalar
constituent $\eta$ is substituted by $\eta^\prime$, the width for
decay $K_{0}^{\ast} \to K \pi$ is not improved but suppressed. In
fact, when $\eta$ meson is substituted by $\eta^\prime$ meson, the
coupling constant $g_{_{K_0^{\ast\pm}}}$ is improved to
$g_{_{K_0^{\ast\pm}}} = 20.54~{\rm GeV}$, but coupling constant
$g_{K^\ast K \eta^\prime}$ is suppressed to $g_{K^\ast K
\eta^\prime} = 1.095$ when the mixing angle $\theta_P =
-9.95^\circ$~\cite{Venugopal:1998fq} and $f_8 = 1.26
f_\pi$~\cite{Feldmann:1998vh} are applied. So that, compared to the
$\eta$ meson case, the partial width $\Gamma(K_{0}^{\ast+} \to K
\pi)$ from the final state interaction is suppressed by a factor
$(20.54\times 1.095)^2/(12.99\times 6.14)^2 \simeq 0.078$. For the
contribution from the contact diagram, explicit calculation shows
that the partial width due to this diagram is about 34\% of that of
the $\eta$ meson case. So that the partial width from the sum of
these two diagrams is about one third of that of the $\eta$ meson
case.

One may notice that another possibility to improve the numerical
value of the width for decay $K_{0}^{\ast} \to K \pi$ is to enlarge
the parameter $\Lambda$ in the form factor $\mathcal{F}(q^2)$. In
fact, we check that to make  the numerical value of the width for
decay $K_{0}^{\ast} \to K \pi$ agrees with the data, $\Lambda \simeq
1.5~$GeV. But in this case, $\Gamma (a_0^+ \to \eta \pi^+) =
749~$MeV and $\Gamma (a_0^0 \to \eta \pi^0) = 665~$MeV. Both of them
are much larger than the data.

At last, we would like to mention that, we have estimated the decay
of $K_0^\ast(1430) \to K \pi$ rudely. In this case, the singlet must
be $\eta^\prime$. Our result for the partial width is $\sim
124.0~$MeV which is about a half of the data and the improvement is
mainly from the phase space. In this sense, it seems that
$\sigma(600), f_0(980), a_(980)$ and $K_0^{\ast}(1430)$ can be
classified into a same nonet. However, this deserves further
systematically study.

In conclusion, our explicit result shows that, in the hadronic
molecular model, it is difficult to arrange the scalar mesons with
masses below 1.0 GeV in the same nonet.

\acknowledgments

\label{ACK}

We would like to thank Profs. Yu-Bing Dong, Qiang Zhao, Bing-Song
Zou from IHEP, Beijing, Prof.Yue-Liang Wu from ITP, Beijing and
Dr.Ping Wang from JLB for the valuable discussions we had with them.
We would also thanks to Prof.J.R.Peaez from Madrid University for
his valuable comments.

\appendix

\section{Integral formulas}

\label{App:int}

In this appendix, we will derive some integral formulae of the
one-loop integrals for $a_0 \to \eta \pi$ decay. To derive the
relations we have used $p=p_1 + p_2$ and $p^2 = m_S^2, p_i^2 =
m_i^2$.

\subsection{Integral formulae for the contact diagram}

In this subsection, we will reduce the momentum integral for the
contact diagram to the standard loop integral function. For
$I_D^{(A)}$, one has
\begin{eqnarray}
I_D^{(A)}(m_S;M_{c_1},M_{c_2}; m_1,m_2) & = & g_{P_1P_2P_3P_4}^{D}\int\frac{d^4k}{(2\pi)^4}\mathcal{F}(k^2)\frac{i}{k^2-M_{c_1}^2}\frac{i}{(k+p)^2-M_{c_2}^2}\nonumber\\
& &  \;\;\;\;\;\;\;\;\;\;\;\; \times \Big[ [(k+p)+p_1]\cdot(k-p_2)+(p_1-k)\cdot[-p_1-(k+p)] \Big]\nonumber\\
& = & - ig_{P_1P_2P_3P_4}^{D}\Big[-( 3m_S^2 - m_1^2 -m_2^2 - \Lambda^2 - M_{c_2}^2)\nonumber\\
& &  \;\;\;\;\;\;\;\;\;\;\;\;\;\;\;\;\;\;\;\;\; \times I_0(m_S;M_{c_1},M_{c_2};\Lambda) \nonumber\\
& & \;\;\;\;\;\;\;\;\;\;\;\;\;\;\;\;\;\;\;\;\;  +
I_2^{(a)}(m_S;M_{c_1},M_{c_2};\Lambda) + I_2^{(b)}(M_{c_1};\Lambda)
\Big]
\end{eqnarray}
where the relevant functions are defined as
\begin{eqnarray}
I_{2}^{(a)}(m_S; M_{c_1},M_{c_2};\Lambda) & = &
\frac{1}{i}\int\frac{d^4k}{(2\pi)^4}{\cal
F}(k^2) \frac{1}{k^2-M_{c_1}^2}\frac{k^2-\Lambda^2}{(k+p)^2-M_{c_2}^2} \nonumber\\
& = &
(M_{c_1}^2-\Lambda^2)\frac{1}{16\pi^2}[B_0(m_S^2,M_{c_1}^2,M_{c_2}^2)-
B_0(m_S^2,\Lambda^2,M_{c_2}^2)] \nonumber\\
I_{2}^{(b)}(M_{c_1};\Lambda) & = &
\frac{1}{i}\int\frac{d^4k}{(2\pi)^4}{\cal
F}(k^2) \frac{1}{k^2-M_{c_1}^2} \nonumber\\
& = &
\frac{1}{16\pi^2}[A_0(M_{c_1}^2)-A_0(\Lambda^2)-(M_{c_1}^2-\Lambda^2)\frac{d}{d\Lambda^2}A_0(\Lambda^2)] \nonumber\\
I_{0}(m_S; M_{c_1},M_{c_2};\Lambda) & = &
\frac{1}{i}\int\frac{d^4k}{(2\pi)^4}{\cal
F}(k^2) \frac{1}{k^2-M_{c_1}^2}\frac{1}{(k+p)^2-M_{c_2}^2}\nonumber\\
& = & \frac{1}{16\pi^2}[B_0(m_S^2,M_{c_1}^2,M_{c_2}^2) - B_0(m_S^2,\Lambda^2,M_{c_2}^2)\nonumber\\
& & \;\;\;\;\;\;\;\;\;\; -(M_{c_1}^2 - \Lambda^2)
\frac{d}{d\Lambda^2}B_0(m_S^2,\Lambda^2,M_{c_2}^2)]
\end{eqnarray}
And for $I_{ND}^{(A)}$ we have
\begin{eqnarray}
I_{ND}^{(A)}(m_S;M_1,M_2) & = &
g_{P_1P_2P_3P_4}^{ND}~\int\frac{d^4k}{(2\pi)^4}\mathcal{F}(k^2)\frac{i}{k^2-m_{c_1}^2}\frac{i}{(k+p)^2-m_{c_2}^2}\nonumber\\
& = & -i g_{P_1P_2P_3P_4}^{ND}~I_0(m_S;M_{c_1},M_{c_2};\Lambda)
\end{eqnarray}

\subsection{Integral formulae for the final state interaction diagram}

In this subsection, we will reduce the momentum integral for the
final state interaction diagram to the standard loop integral
function. After some algebra, one has
\begin{eqnarray}
I^{(B)} (m_S;M_{c_1},M_{c_2};m_V; m_1,m_2) & = & \int\frac{d^4k}{(2\pi)^4}{\cal F}(k^2)\frac{i}{(k-p_1)^2-m_{c_1}^2}\frac{i}{(k+p_2)^2-m_{c_2}^2}\frac{-i}{k^2-m_{V}^2}\nonumber\\
& & \times (2p_1-k)_\mu [g_{\mu\nu} - \frac{k_\mu
k_\nu}{m_{K^\ast}^2}](k+2p_2)_\nu \nonumber\\
& = & I_{0}^{(a)}(m_S;M_{c_1},M_{c_2};m_V;m_1,m_2;\Lambda) +
\frac{1}{m_V^2}
I_{2}^{(a)}(m_V;\Lambda) \nonumber\\
& & - [1 - \frac{1}{m_V^2}(M_{c_2}^2 - m_2^2)]
I_{0}^{(b)}(M_{c_2};m_V;m_2;\Lambda) \nonumber\\
& & - [1 - \frac{1}{m_V^2}(M_{c_1}^2 - m_1^2)]
I_{0}^{(c)}(M_{c_1},m_V;m_1;\Lambda)\nonumber\\
& & - \Big[M_{c_1}^2 + M_{c_2}^2 - m_V^2 - 2m_S^2 + m_1^2 + m_2^2 \nonumber\\
& & \;\;\;\;\; - \frac{1}{m_V^2}(M_{c_1}^2 - m_1^2)(M_{c_2}^2 -
m_2^2)\Big]\nonumber\\
& & \times I_{-2}(m_S;M_{c_1},M_{c_2};m_V;m_1,m_2;\Lambda)
\end{eqnarray}
where
\begin{eqnarray}
I_{0}^{(a)}(m_S; M_{c_1},M_{c_2}; m_V; m_1,m_2;\Lambda) & = &
\frac{1}{i}\int\frac{d^4k}{(2\pi)^4}{\cal
F}(k^2)\nonumber\\
& & \times \frac{1}{(k-p_1)^2-M_{c_1}^2}\frac{1}{(k+p_2)^2-M_{c_2}^2}\frac{k^2-\Lambda^2}{k^2-m_V^2} \nonumber\\
& = & (m_V^2-\Lambda^2)\frac{1}{16\pi^2}[C_0(m_1^2,m_2^2,m_S^2,M_{c_1}^2,m_V^2,m_{c_2}^2)\nonumber\\
& & \;\;\;\;\;\;\;\;\;\;\;\;\;\;\;\;\;\;\;\;\;\;\;\;\;\;\;\; -
C_0(m_1^2,m_2^2,m_S^2,M_{c_1}^2,\Lambda^2,m_{c_2}^2)] \nonumber\\
I_{0}^{(b)}(M_{c_2}; m_V;m_2;\Lambda) & = &
\frac{1}{i}\int\frac{d^4k}{(2\pi)^4}{\cal F}(k^2) \frac{1}{(k+p_2)^2-M_{c_2}^2}\frac{1}{k^2-m_V^2} \nonumber\\
& = & \frac{1}{16\pi^2}[B_0(m_2^2,m_V^2,m_{c_2}^2) - B_0(m_2^2,\Lambda^2,m_{c_2}^2)\nonumber\\
& & \;\;\;\;\;\;\;\;\;\; -(m_V^2 - \Lambda^2) \frac{d}{d\Lambda^2}B_0(m_2^2,\Lambda^2,m_{c_2}^2)] \nonumber\\
I_{0}^{(c)}(M_{c_1}; m_V;m_1;\Lambda) & = &
\frac{1}{i}\int\frac{d^4k}{(2\pi)^4}{\cal F}(k^2,\Lambda^2) \frac{1}{(k+p_1)^2-M_{c_1}^2}\frac{1}{k^2-m_V^2} \nonumber\\
& = & \frac{1}{16\pi^2}[B_0(m_1^2,m_V^2,m_{c_1}^2) - B_0(m_1^2,\Lambda^2,m_{c_1}^2)\nonumber\\
& & \;\;\;\;\;\;\;\;\;\; -(m_V^2 - \Lambda^2) \frac{d}{d\Lambda^2}B_0(m_1^2,\Lambda^2,m_{c_1}^2)] \nonumber\\
I_{2}^{(a)}(m_V; \Lambda) & = &
\frac{1}{i}\int\frac{d^4k}{(2\pi)^4}{\cal
F}(k^2) \frac{1}{k^2-m_V^2} \nonumber\\
& = & \frac{1}{16\pi^2}[A_0(m_V^2) - A_0(\Lambda^2) -(m_V^2 - \Lambda^2) \frac{d}{d\Lambda^2}A_0(\Lambda^2)] \nonumber\\
I_{-2}(m_S; M_{c_1},M_{c_2}; m_V; m_1,m_2;\Lambda) & = &
\frac{1}{i}\int\frac{d^4k}{(2\pi)^4}{\cal
F}(k^2)\nonumber\\
& & \times \frac{1}{(k-p_1)^2-M_{c_1}^2}\frac{1}{(k+p_2)^2-M_{c_2}^2}\frac{1}{k^2-m_V^2} \nonumber\\
& = & \frac{1}{16\pi^2}[C_0(m_1^2,m_2^2,m_S^2,M_{c_1}^2,m_V^2,m_{c_2}^2)\nonumber\\
& & \;\;\;\;\;\;\;\;\;\; - C_0(m_1^2,m_2^2,m_S^2,M_{c_1}^2,\Lambda^2,m_{c_2}^2)\nonumber\\
& & \;\;\;\;\;\;\;\;\;\; -(m_V^2 - \Lambda^2)
\frac{d}{d\Lambda^2}C_0(m_1^2,m_2^2,m_S^2,M_{c_1}^2,\Lambda^2,m_{c_2}^2)]
\end{eqnarray}



\begin{thebibliography}{99}


\bibitem{Amsler:2008zz}
  C.~Amsler {\it et al.}  [Particle Data Group],
  Phys.\ Lett.\  B {\bf 667}, 1 (2008).

\bibitem{Godfrey:1998pd}
  S.~Godfrey and J.~Napolitano,
  Rev.\ Mod.\ Phys.\  {\bf 71}, 1411 (1999)
  [arXiv:hep-ph/9811410].

\bibitem{Close:2002zu}
  F.~E.~Close and N.~A.~Tornqvist,
  J.\ Phys.\ G {\bf 28}, R249 (2002)
  [arXiv:hep-ph/0204205].

\bibitem{Jaffe:1976ig}
  R.~L.~Jaffe,
  Phys.\ Rev.\  D {\bf 15}, 267 (1977).

\bibitem{Jaffe:1976ih}
  R.~L.~Jaffe,
  Phys.\ Rev.\  D {\bf 15}, 281 (1977).




\bibitem{Alford:2000mm}
  M.~G.~Alford and R.~L.~Jaffe,
  Nucl.\ Phys.\  B {\bf 578}, 367 (2000)
  [arXiv:hep-lat/0001023].

\bibitem{Maiani:2004uc}
  L.~Maiani, F.~Piccinini, A.~D.~Polosa and V.~Riquer,
  Phys.\ Rev.\ Lett.\  {\bf 93}, 212002 (2004)
  [arXiv:hep-ph/0407017].


\bibitem{Oller:1998hw}
  J.~A.~Oller, E.~Oset and J.~R.~Pelaez,
  Phys.\ Rev.\  D {\bf 59}, 074001 (1999)
  [Erratum-ibid.\  D {\bf 60}, 099906 (1999\ ERRAT,D75,099903.2007)]
  [arXiv:hep-ph/9804209].


\bibitem{Ishida:1999qk}
  M.~Ishida,
  Prog.\ Theor.\ Phys.\  {\bf 101}, 661 (1999)
  [arXiv:hep-ph/9902260].

\bibitem{Scadron:2003yg}
  M.~D.~Scadron, G.~Rupp, F.~Kleefeld and E.~van Beveren,
  Phys.\ Rev.\  D {\bf 69}, 014010 (2004)
  [Erratum-ibid.\  D {\bf 69}, 059901 (2004)]
  [arXiv:hep-ph/0309109].


\bibitem{Tornqvist:1999tn}
  N.~A.~Tornqvist,
  Eur.\ Phys.\ J.\  C {\bf 11}, 359 (1999)
  [arXiv:hep-ph/9905282].



\bibitem{Tornqvist:1995kr}
  N.~A.~Tornqvist,
  Z.\ Phys.\  C {\bf 68}, 647 (1995)
  [arXiv:hep-ph/9504372].


\bibitem{Weinstein:1990gu}
  J.~D.~Weinstein and N.~Isgur,
  Phys.\ Rev.\  D {\bf 41}, 2236 (1990).


\bibitem{Pelaez:2004xp}
  J.~R.~Pelaez,
  Mod.\ Phys.\ Lett.\  A {\bf 19}, 2879 (2004)
  [arXiv:hep-ph/0411107].

\bibitem{Pelaez:2003dy}
  J.~R.~Pelaez,
  Phys.\ Rev.\ Lett.\  {\bf 92}, 102001 (2004)
  [arXiv:hep-ph/0309292].

\bibitem{Pelaez:2006nj}
  J.~R.~Pelaez and G.~Rios,
  Phys.\ Rev.\ Lett.\  {\bf 97}, 242002 (2006)
  [arXiv:hep-ph/0610397].





\bibitem{Giacosa:2007bs}
  F.~Giacosa, T.~Gutsche and V.~E.~Lyubovitskij,
  Phys.\ Rev.\  D {\bf 77}, 034007 (2008)
  [arXiv:0710.3403 [hep-ph]].



\bibitem{Branz:2007xp}
  T.~Branz, T.~Gutsche and V.~E.~Lyubovitskij,
  Eur.\ Phys.\ J.\  A {\bf 37}, 303 (2008)
  [arXiv:0712.0354 [hep-ph]].



\bibitem{Branz:2008ha}
  T.~Branz, T.~Gutsche and V.~E.~Lyubovitskij,
  Phys.\ Rev.\  D {\bf 78}, 114004 (2008)
  [arXiv:0808.0705 [hep-ph]].


\bibitem{Branz:2008qm}
  T.~Branz, T.~Gutsche and V.~E.~Lyubovitskij,
  AIP Conf.\ Proc.\  {\bf 1030}, 118 (2008)
  [arXiv:0805.1647 [hep-ph]].


\bibitem{Branz:2008cb}
  T.~Branz, T.~Gutsche and V.~E.~Lyubovitskij,
  arXiv:0812.0942 [hep-ph].





\bibitem{Faessler:2003yf}
  A.~Faessler, T.~Gutsche, M.~A.~Ivanov, V.~E.~Lyubovitskij and P.~Wang,
  Phys.\ Rev.\  D {\bf 68}, 014011 (2003)
  [arXiv:hep-ph/0304031].

\bibitem{vanBeveren:2001kf}
  E.~van Beveren and G.~Rupp,
  Eur.\ Phys.\ J.\  C {\bf 22}, 493 (2001)
  [arXiv:hep-ex/0106077].



\bibitem{Celenza:2000uk}
  L.~S.~Celenza, S.~f.~Gao, B.~Huang, H.~Wang and C.~M.~Shakin,
  Phys.\ Rev.\  C {\bf 61}, 035201 (2000).

\bibitem{Dai:2003ip}
  Y.~B.~Dai and Y.~L.~Wu,
  Eur.\ Phys.\ J.\  C {\bf 39}, S1 (2005)
  [arXiv:hep-ph/0304075].


\bibitem{:2008my}
  Bhavyashri, K.~B.~Vijaya Kumar, Y.~L.~Ma and A.~Prakash,
  arXiv:0811.4308 [hep-ph].


\bibitem{Cheng:2009dg}
  H.~Y.~Cheng,
  arXiv:0901.0741 [hep-ph].


\bibitem{Faessler:2007gv}
  A.~Faessler, T.~Gutsche, V.~E.~Lyubovitskij and Y.~L.~Ma,
  Phys.\ Rev.\  D {\bf 76}, 014005 (2007)
  [arXiv:0705.0254 [hep-ph]].



\bibitem{Faessler:2007us}
  A.~Faessler, T.~Gutsche, V.~E.~Lyubovitskij and Y.~L.~Ma,
  Phys.\ Rev.\  D {\bf 76}, 114008 (2007)
  [arXiv:0709.3946 [hep-ph]].


\bibitem{Dong:2008gb}
  Y.~b.~Dong, A.~Faessler, T.~Gutsche and V.~E.~Lyubovitskij,
  Phys.\ Rev.\  D {\bf 77}, 094013 (2008)
  [arXiv:0802.3610 [hep-ph]].



\bibitem{Faessler:2008vc}
  A.~Faessler, T.~Gutsche, V.~E.~Lyubovitskij and Y.~L.~Ma,
  arXiv:0801.2232 [hep-ph].

\bibitem{Ma:2008hc}
  Y.~L.~Ma,
  J.\ Phys.\ G {\bf 36}, 055004 (2009)
  [arXiv:0808.3764 [hep-ph]].


\bibitem{Napsuciale:2004au}
  M.~Napsuciale and S.~Rodriguez,
  Phys.\ Rev.\  D {\bf 70}, 094043 (2004)
  [arXiv:hep-ph/0407037].


\bibitem{Wu:2007jh}
  J.~J.~Wu, Q.~Zhao and B.~S.~Zou,
  Phys.\ Rev.\  D {\bf 75}, 114012 (2007)
  [arXiv:0704.3652 [hep-ph]].

\bibitem{Wu:2008hx}
  J.~J.~Wu and B.~S.~Zou,
  Phys.\ Rev.\  D {\bf 78}, 074017 (2008)
  [arXiv:0808.2683 [hep-ph]].


\bibitem{HerreraSiklody:1996pm}
  P.~Herrera-Siklody, J.~I.~Latorre, P.~Pascual and J.~Taron,
  Nucl.\ Phys.\  B {\bf 497}, 345 (1997)
  [arXiv:hep-ph/9610549].



\bibitem{Harada:1995sj}
  M.~Harada and J.~Schechter,
  Phys.\ Rev.\  D {\bf 54}, 3394 (1996)
  [arXiv:hep-ph/9506473].


\bibitem{Gasser:1983yg}
  J.~Gasser and H.~Leutwyler,
  Annals Phys.\  {\bf 158}, 142 (1984).

\bibitem{Gasser:1984gg}
  J.~Gasser and H.~Leutwyler,
  Nucl.\ Phys.\  B {\bf 250}, 465 (1985).


\bibitem{Machleidt:1987hj}
  R.~Machleidt, K.~Holinde and C.~Elster,
  Phys.\ Rept.\  {\bf 149} (1987) 1.



\bibitem{Hahn:1998yk}
  T.~Hahn and M.~Perez-Victoria,
  Comput.\ Phys.\ Commun.\  {\bf 118}, 153 (1999)
  [arXiv:hep-ph/9807565].




\bibitem{Giacosa:2006rg}
  F.~Giacosa,
  Phys.\ Rev.\  D {\bf 74}, 014028 (2006)
  [arXiv:hep-ph/0605191].


\bibitem{Palomar:2003rb}
  J.~E.~Palomar, L.~Roca, E.~Oset and M.~J.~Vicente Vacas,
  Nucl.\ Phys.\  A {\bf 729}, 743 (2003)
  [arXiv:hep-ph/0306249].



\bibitem{Harada:2003jx}
  M.~Harada and K.~Yamawaki,
  Phys.\ Rept.\  {\bf 381} (2003) 1
  [arXiv:hep-ph/0302103].



\bibitem{Urech:1995ry}
  R.~Urech,
  Phys.\ Lett.\  B {\bf 355} (1995) 308
  [arXiv:hep-ph/9504238].

\bibitem{Bramon:1997va}
  A.~Bramon, R.~Escribano and M.~D.~Scadron,
  Eur.\ Phys.\ J.\  C {\bf 7}, 271 (1999)
  [arXiv:hep-ph/9711229].


\bibitem{Achasov:2000ym}
  M.~N.~Achasov {\it et al.},
  Phys.\ Lett.\  B {\bf 485}, 349 (2000)
  [arXiv:hep-ex/0005017].


\bibitem{Achasov:2000ku}
  M.~N.~Achasov {\it et al.},
  Phys.\ Lett.\  B {\bf 479}, 53 (2000)
  [arXiv:hep-ex/0003031].


\bibitem{Aloisio:2002bt}
  A.~Aloisio {\it et al.}  [KLOE Collaboration],
  Phys.\ Lett.\  B {\bf 537}, 21 (2002)
  [arXiv:hep-ex/0204013].

\bibitem{Aloisio:2002bsa}
  A.~Aloisio {\it et al.}  [KLOE Collaboration],
  Phys.\ Lett.\  B {\bf 536}, 209 (2002)
  [arXiv:hep-ex/0204012].


\bibitem{Akhmetshin:1999di}
  R.~R.~Akhmetshin {\it et al.}  [CMD-2 Collaboration],
  Phys.\ Lett.\  B {\bf 462}, 380 (1999)
  [arXiv:hep-ex/9907006].

\bibitem{Venugopal:1998fq}
  E.~P.~Venugopal and B.~R.~Holstein,
  Phys.\ Rev.\  D {\bf 57}, 4397 (1998)
  [arXiv:hep-ph/9710382].

\bibitem{Feldmann:1998vh}
  T.~Feldmann, P.~Kroll and B.~Stech,
  Phys.\ Rev.\  D {\bf 58}, 114006 (1998)
  [arXiv:hep-ph/9802409].





















\end{thebibliography}
\end{document}